\documentclass[12pt,preprint]{aastex}
\shorttitle{}
\shortauthors{Nesvorn\'y}
\begin{document}
\title{The Evidence for Slow Migration of Neptune from the Inclination Distribution 
of Kuiper Belt Objects}
\author{David Nesvorn\'y}
\affil{Department of Space Studies, Southwest Research Institute, 1050 Walnut St., \\
Suite 300, Boulder, CO 80302, USA} 
\begin{abstract}
Much of the dynamical structure of the Kuiper belt can be explained if Neptune migrated over several AU, 
and/or if Neptune was scattered to an eccentric orbit during planetary 
instability. An outstanding problem with the existing formation models is that the distribution of 
orbital inclinations they predicted is narrower than the one inferred from observations. Here 
we perform numerical simulations of Kuiper belt formation starting from an initial state with 
Neptune at $20<a_{\rm N,0}<30$ AU and a dynamically cold outer disk extending from beyond $a_{\rm N,0}$ 
to 30 AU. Neptune's orbit is migrated into the disk on an e-folding timescale $1 \leq \tau \leq 
100$ Myr. A small fraction ($\sim$$10^{-3}$) of the disk planetesimals become implanted into the Kuiper 
belt in the simulations. By analyzing the orbital distribution of the implanted bodies in different 
cases we find that the inclination constraint implies that $\tau\gtrsim10$ Myr and $a_{\rm N,0}\lesssim25$ AU. 
The models with $\tau<10$~Myr do not satisfy the inclination constraint, because there is 
not enough time for various dynamical processes to raise inclinations. The slow migration of Neptune 
is consistent with other Kuiper belt constraints, and with recently developed models of planetary 
instability/migration. Neptune's eccentricity and inclination are never large in these models 
($e_{\rm N}<0.1$, $i_{\rm N}<2^\circ$), as required to avoid excessive orbital excitation in the 
$>$40 AU region, where the Cold Classicals presumably formed. 
\end{abstract}
\section{Background}
\begin{figure}
\epsscale{0.6}
\plotone{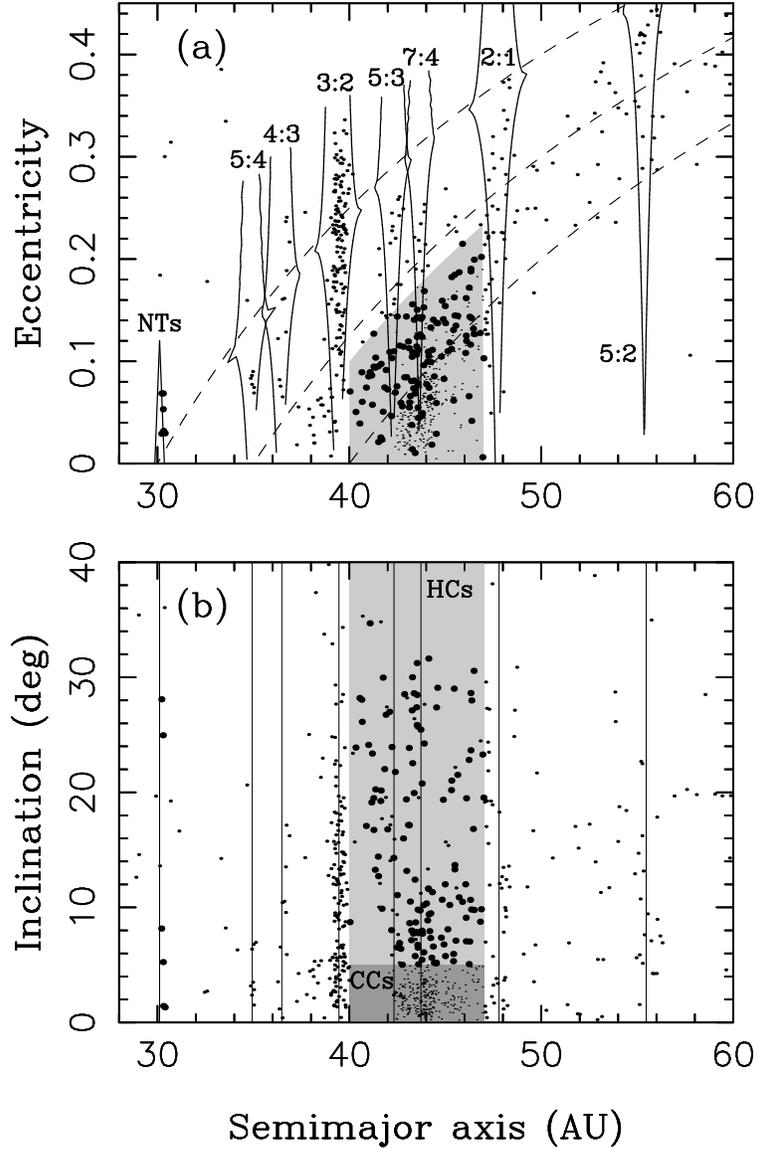}
\caption{The orbital elements of KBOs observed in three or more oppositions. Various dynamical classes 
are highlighted. The HCs with $i>5^\circ$ and NTs are denoted by larger dots, and the CCs are denoted 
by smaller dots. Note the wide inclination 
distribution of the HCs in panel (b) with inclinations reaching above $\simeq30^\circ$. The solid lines 
in panel (a) follow the borders of important mean motion resonances. For the NTs, we show the approximate 
location of stable librations from Nesvorn\'y \& Dones (2002). The low-inclination orbits
with $40<a<42$ AU are unstable due to secular resonance overlap ($\nu_7$ and $\nu_{8}$; 
Kn\v{e}\v{z}evi\'c 1991, Duncan et al. 1995).}
\label{fig1}
\end{figure}

The Kuiper belt is a diverse population of trans-Neptunian bodies (Figure \ref{fig1}). Based on dynamical 
considerations, the Kuiper Belt Objects (KBOs) are classified into several groups: the resonant 
populations, classical belt, scattered/scattering disk, and detached objects (also known 
as the fossilized scattered disk). See Gladman et al. (2008) for a formal definition of these groups. 
The resonant populations are a fascinating feature of the Kuiper belt. They give the Kuiper belt an 
appearance of a bar code with individual bars centered at the resonant orbital periods. While Pluto and 
Plutinos in the 3:2 resonance with Neptune (orbital period $\simeq$250 years) are the largest and the 
best-characterized resonant group, nearly every resonance hosts a large population of bodies.
The resonant bodies are long-lived, because they are phase-protected by the resonance from close encounters 
with Neptune. The orbits of the scattered/scattering disk objects, on the other hand, evolved and keep 
evolving by close encounters with Neptune. These objects tend to have long orbital periods and to be
detected near their orbital perihelion when the heliocentric distance is $\sim$30~AU. Their neighbors, 
the detached objects, have a slightly larger perihelion distance than the scattered/scattering objects 
and semimajor axes beyond the 2:1 resonance ($a>47.8$~AU). The detached objects probably suffered close 
encounters with Neptune in the past, were scattered to orbits with large semimajor axes and eccentricities
($e>0.24$ defines them in Gladman et al. 2008), but then they became ``detached'' from Neptune when some 
process increased their perihelion distance (or when Neptune's orbit circularized; Levison et al. 2008).

The classical belt is a population of trans-Neptunian bodies dynamically defined as having non-resonant 
orbits with perihelion distances that are large enough to avoid close encounters with Neptune. They can 
be thought as of being related to the detached objects but having orbits with modest orbital eccentricities 
($e<0.24$ according to Gladman et al. 2008). Here we consider the main classical belt located between 
the 3:2 and 2:1 resonances with Neptune ($39.4<a<47.8$ AU), because this is where most known classical objects 
reside. It is useful to divide the main belt into the dynamically ``cold'' and ``hot'' components, mainly 
because the inclination distribution of the main belt orbits is bimodal (Brown 2001), 
hinting at different dynamical origins for these components. Here we adopt an 
approximate separation, with Cold Classicals (CCs) being defined as having $i<5^\circ$ and Hot 
Classicals (HCs) as $i>5^\circ$. Note that this definition is somewhat arbitrary, because the continuous 
inclination distribution near $i=5^\circ$ indicates that mixing between the two components 
must have happened (e.g., Morbidelli et al. 2008, Volk \& Malhotra 2011).

While the HCs share many similarities with other dynamical classes of KBOs (e.g., scattered disk, Plutinos), 
the CCs have several unique properties. Specifically, (1) the CCs have distinctly red colors (e.g., Tegler \& 
Romanishin 2000) that may have resulted from space weathering of surface ices, such as ammonia 
(Brown et al. 2011), that are stable beyond $\sim$35 AU. (2) A large fraction of the 100-km-class CCs are wide 
binaries with nearly equal size components (Noll et al. 2008a,b). (3) The albedos of the CCs are generally higher 
than those of the HCs (Brucker et al. 2009). And finally, (4) the size distribution of the CCs is markedly different 
from those of the hot and scattered populations, in that it shows a very steep slope at large sizes (e.g., 
Bernstein et al. 2004, Fraser et al. 2014), and lacks very large objects (Levison \& Stern 2001). 
The most straightforward interpretation of these properties is that the CCs formed and/or dynamically 
evolved by different processes than other trans-Neptunian populations.

Following the pioneering work of Malhotra (1993, 1995), studies of Kuiper belt dynamics first considered 
the effects of outward migration of Neptune that can explain the prominent populations of KBOs in 
major resonances (Hahn \& Malhotra 1999, 2005; Chiang \& Jordan 2002; Chiang et al. 2003; Levison \& Morbidelli 2003;
Gomes 2003; Murray-Clay \& Chiang 2005, 2006). With the advent of the notion that the early solar system may 
have suffered a dynamical instability (Thommes et al. 1999, Tsiganis et al. 2005), the focus broadened, with 
the more recent theories invoking a transient phase with an eccentric orbit of Neptune (Levison et al. 2008, 
Morbidelli et al. 2008, Batygin et al. 2011, Wolff~et~al.~2012, Dawson \& Murray-Clay 2012). 

The emerging consensus is that the HCs, together with the resonant, scattered and detached populations, 
formed in a massive planetesimal disk at $\lesssim$30 AU, and were dynamically scattered onto their current 
orbits by migrating (and possibly eccentric) Neptune, while the CCs formed at $>$40~AU and survived Neptune's 
early ``wild days'' relatively unharmed (Batygin et al. 2011, Wolff et al. 2012). 
The main support for this model comes from the unique properties of the CCs, which would be difficult to explain if 
the HCs and CCs had similar formation 
locations (and dynamical histories). For example, the wide binaries observed among the CCs would not survive 
scattering encounters with Neptune (Parker \& Kavelaars 2010). Moreover, if the CCs evolved from the high-eccentricity 
Neptune-crossing orbits, this process should produce a gradient in $e$ with more orbits having large $e$ 
and fewer orbits having small $e$. The CCs do not show such a trend. Instead, low eccentricities prevail in 
that population.
\section{The Inclination Problem}
The inclination distribution of various populations in the Kuiper belt can be represented by 
$N(i)\,{\rm d}i = \sin i \exp(-i^2/2\sigma_i^2)\,{\rm d}i$, where $\sigma_i$ is a parameter (Brown 2001). 
In the main belt, the inclination distribution is bimodal and two components are needed: $\sigma_i\simeq2.0^\circ$ 
for the low-$i$ CCs and $\sigma_i =8^\circ$-$17^\circ$ for the high-$i$ HCs (Brown 2001, Kavelaars et al. 2008, 
2009, Gulbis et al. 2010). The low inclinations of CCs are in line with the expectation that they 
formed from a dynamically cold disk at $>$40 AU, and their orbits were never excited too much by 
subsequent dynamical processes. The high inclinations of the HCs, on the other hand, are more challenging 
to explain (see below). Moreover, there is some evidence from high-latitude surveys that 
$\sin i \exp(-i^2/2\sigma_i^2)$ may be somewhat inadequate, because the drop-off at large values of $i$ is 
probably steeper than expected from this functional dependence (Petit et al. 2015). For this reason, it is 
possible that $N(i)\,{\rm d}i = \sin i \exp(-(i-i_0)^2/\sigma_i^2)\,{\rm d}i$ with $i_0 \gtrsim 5^\circ$ may 
better represent the underlying distribution.

The HC distribution with relatively high orbital inclinations is shared among several other Kuiper belt populations 
as well, including Neptune Trojans (NTs), and the resonant and scattered objects. Eight NTs 
are currently known. Four of them have orbital inclinations $i<10^\circ$, and four have $25<i<30^\circ$. 
This could mean that the distribution is bimodal, but Parker (2015) showed that the bimodality of the 
underlying inclination distribution cannot be demonstrated with confidence from the existing data. If the 
distribution is parametrized by a single term, $N(i)\,{\rm d}i = \sin i \exp(-i^2/2\sigma_i^2)\,{\rm d}i$, 
the NTs are inferred to have $\sigma_i>11^\circ$ with a 95\% confidence (Parker 2015). Plutinos in 
the 3:2 resonance with Neptune are well represented by a single term with $\sigma_i\simeq11^\circ$ according 
to Gulbis et al. (2010), or $\sigma_i\simeq15^\circ$ according to Kavelaars et al. (2008) and Gladman et al. 
(2012). Interestingly, the CC-like component with low orbital inclinations is not found in the 
Plutino population.   

The wide inclination distribution of the HCs, NTs, Plutinos and other resonant populations poses an 
important constraint on dynamical models of Kuiper belt formation.\footnote{Assuming the current 
configuration of planets, long-term orbital dynamics in the Kuiper belt region cannot explain the 
high inclinations of the KBOs (Kuchner et al. 2002, Li et al. 2014).} It implies that some dynamical process 
must have increased the inclinations by 10-15$^\circ$ on average, and by $\simeq$30$^\circ$ at least in some cases.
For example, the inclination constraint can be used to rule out a model in which these populations arise from 
a dynamically cold planetesimal disk at $>$30 AU, simply because the orbital inclinations are not excited 
in this region during Neptune's migration (passing mean motion resonances do not affect inclinations much).
Hahn \& Malhotra (2005) investigated this issue in detail and found that starting with moderately excited orbits 
of planetesimals (e.g., $i<10^\circ$) does not resolve the problem, because the final inclination distribution 
is still not wide enough. The only way to make things work in the context of the Hahn \& Malhotra model would 
be to assume that the inclinations were already high {\it before} Neptune's migration, but that seems 
unsatisfactory, because it is not clear how the inclinations could have been excited beforehand.  
Lykawka \& Mukai (2008) considered dynamical effects of an additional planet in the trans-Neptunian 
region. They found that this putative planet could help to excite inclinations, but the inclination
distribution obtained in their model was not wide enough to match observations well. It lacked orbits 
in the classical belt with $i>15^\circ$, while these orbits are in fact common.

Given the difficulties described above, various theoretical models considered the formation of the HCs and resonant 
populations from a massive planetesimal disk at $<$30 AU (the outer edge of the massive disk is constrained to 
$\simeq$30 AU by Neptune's present orbit; Gomes et al. 2004). To reach $>$40 AU, planetesimals must be 
radially displaced. Levison \& Morbidelli (2003) considered a scenario in which objects were pushed out by the 
2:1 resonance with Neptune. This could work only if Neptune was initially inside $\simeq$19 AU, such that the 
2:1 resonance fell inside the massive disk's outer boundary at $\simeq$30 AU. Gomes (2003), on the other 
hand, suggested that bodies were first scattered to $>$40 AU by having close encounters with Neptune, and 
became dynamically decoupled from Neptune while Neptune was still migrating. If so, the HCs and resonant 
populations would be close relatives of the scattered disk objects. The exact nature of the decoupling process 
for the HCs is uncertain (Gomes 2003, Levison et al. 2008, Dawson \& Murray-Clay 2012), but recent work suggests 
that capture into mean motion resonances with Neptune (2:1, 5:3, 7:4, etc.), and secular/Kozai cycles inside
the resonances may have played an important role (Brasil et al. 2014b).

These results could help to resolve the inclination problem discussed above, because the orbital inclinations 
can be excited when bodies undergo a series of close encounters with Neptune (Gomes 2003). In addition, inclinations 
are increased for orbits that suffer Kozai cycles, because to decouple from Neptune, the eccentricity must  
drop, and the inclination would therefore rise (due to the anticorrelated behavior of $e$ and $i$ caused by the 
Kozai cycles; Kozai 1962). It remains to be shown, however, how these processes operated to affect the Kuiper 
belt, and how the early evolution of planetary orbits is constrained by the dynamical structure of the Kuiper 
belt.  

\begin{figure}
\epsscale{0.7}
\plotone{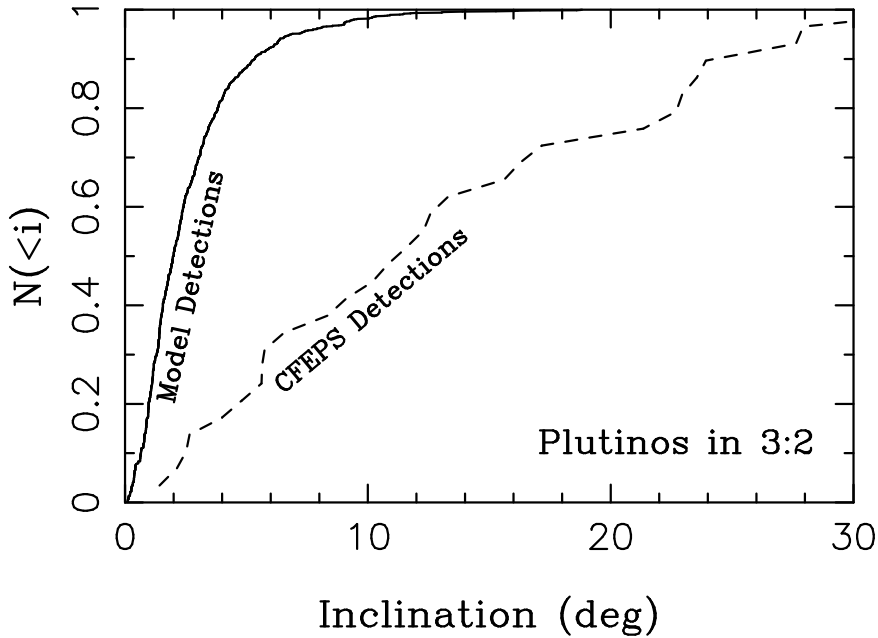}
\caption{The cumulative inclination distribution of Plutinos in the 3:2 resonance with Neptune. The
distribution of 29 Plutinos detected by CFEPS (dashed line, Petit et al. 2011) is 
compared to a model distribution (solid line). The K-S test applied to these distributions
shows that the likelihood that they can be obtained from the same underlying distribution is 
$2\times10^{-13}$. This rules out the model. The model distribution is a result of a numerical simulation where 
we considered Neptune's migration into a dynamically cold disk ($\sigma_e=0.1$ and $\sigma_i=2^\circ$) at 
$a<30$ AU. See Section 3 for a description of the model. Here we used $a_{\rm N,0}=28$ AU, $e_{\rm N,0}=0.1$ and
$i_{\rm N,0}=0.67^\circ$. The radial migration and eccentricity damping were applied to Neptune's orbit 
on an e-folding timescale $\tau=1$ Myr. To compare apples with apples, the CFEPS simulator 
(see Section 4) was used to compute the detection statistics from the population of bodies that survived 
in the 3:2 resonance at the end of the simulation. Thus, both distributions shown in this plot include 
the observational bias of CFEPS. The inclination distribution obtained for larger values of 
$e_{\rm N,0}$ (Levison et al. 2008 used $e_{\rm N,0}=0.3$) is similar to the one shown here for
$e_{\rm N,0}=0.1$.}
\label{iprob}
\end{figure}

As demonstrated in Figure \ref{iprob}, the inclination problem is {\it not} resolved by simply postulating 
that much of the Kuiper belt has been implanted from $<$30 AU (e.g., see discussion in Petit et al. 2011). 
Here we choose to illustrate the inclination
problem with Plutinos, because the 3:2 resonance population is characterized much better than any other resonant 
population (e.g.,  Gladman et al. 2012). Also, Plutinos do not show the bimodal inclination distribution of the 
classical belt, so we do not need to worry about the overlap of different groups. The parameters 
of the numerical model from Figure \ref{iprob} are similar to those used in Levison et al. (2008). We used a 
fast migration regime with an exponential e-folding timescale $\tau=1$ Myr (see next section for our model 
description) to illustrate that such a fast migration of Neptune leads to an implausible result. There is simply 
not enough time in this case to substantially raise the orbital inclinations. 

Here we perform numerical simulations to investigate the inclination problem in detail. Our method and constraints 
are described in Sections 3 and 4, and the results are presented in Section~5. We find that bodies starting with
$a<30$ AU can be implanted into the Kuiper belt by first being scattered by Neptune to $>30$ AU, and then decoupling 
from Neptune by various resonant effects while Neptune is still migrating (see Section 5.2). We show that 
the inclination constraint implies a prolonged phase during which Neptune {\it slowly} migrated (e-folding migration 
timescale $\tau\gtrsim10$ Myr) before reaching its current orbit at $a=30.1$~AU.\footnote{A correlation
between the width of the inclination distribution and the migration timescale/time of capture were 
previously reported by Malhotra (1995) and Levison et al. (2009).} The main effect of slow 
migration is that orbits are allowed to decouple from Neptune relatively late during the migration process. 
Consequently, Neptune is given more time to act, via scattering encounters, on the source population, thus 
increasing the orbital inclinations of bodies before they are implanted into the Kuiper belt. The model with 
Neptune's slow migration is also consistent with other Kuiper belt constraints (Section~5). 
Various implications of this result are discussed in Section 6.   
\section{The Integration Method}
Our numerical integrations track the orbits of four planets (Jupiter to Neptune) and a large number 
of test particles representing the outer planetesimal disk. To set up an integration, Jupiter, Saturn and Uranus 
were placed on their current orbits.\footnote{The dependence of the results on the orbital behavior of Jupiter, 
Saturn and Uranus was found to be minor. We determined this by comparing our nominal results with fixed orbits 
to those obtained when these planets were forced to radially migrate.} Neptune was placed on an orbit 
with semimajor axis $a_{\rm N,0}$, eccentricity $e_{\rm N,0}$, and inclination $i_{\rm N,0}$. To cover the parameter 
space, we set $a_{\rm N,0}=22$, 24, 26 or 28 AU, $e_{\rm N,0}=0$, 0.1 or 0.3, and $i=0^\circ$ or $5^\circ$. We tested 
many different combinations of these parameters to understand their role in Kuiper belt formation. The cases 
with $a_{\rm N,0}=22$ AU or 24 AU, $e_{\rm N,0}=0$ and $i=0^\circ$ are a good proxy for the initial conditions of 
Hahn \& Malhotra (2005), who studied a long-range migration of Neptune, and for the instability/migration models 
developed in Nesvorn\'y \& Morbidelli (2012). The case with $a_{\rm N,0}=28$~AU and $e_{\rm N,0}=0.3$ is similar 
to runs A, B and C in Levison et al. (2008), who motivated their choice by the strong planetary instability 
occurring in the Nice model (Tsiganis et al. 2005, Morbidelli et al. 2007). The cases with $e_{\rm N,0}\lesssim0.1$ 
were favored by Dawson \& Murray-Clay (2012) from the CC-related constraints.  

The {\tt swift\_rmvs4} code (Levison \& Duncan 1994) was used to follow the evolution of planets and disk 
particles. The {\tt swift\_rmvs4} code was modified to include fictitious forces that mimic Neptune's radial 
migration and damping. These forces were parametrized by exponential e-folding timescales $\tau_a$, $\tau_e$ 
and $\tau_i$, where $\tau_a$ controls the radial migration rate, and $\tau_e$ and $\tau_i$ control the damping 
rates of $e$ and $i$. Here we set $\tau_a \sim \tau_e \sim \tau_i$ $(=\tau)$, because such roughly comparable 
timescales were suggested by previous work. Specifically, we used $\tau=1$, 3, 10, 30 and 100 Myr, where 
$\tau=1$ Myr corresponds to the case considered by Levison et al. (2008), while $\tau \gtrsim 10$~Myr is preferred 
from the instability simulations of Nesvorn\'y \& Morbidelli (2012). 
By fine tuning the migration parameters, the final semimajor axis of Neptune was set to be within 0.05 AU of its 
current mean $a_{\rm N}=30.11$~AU, and the orbital period ratio, $P_{\rm N}/P_{\rm U}$, where $P_{\rm N}$ and 
$P_{\rm U}$ are the orbital periods of Neptune and Uranus, was adjusted to end up within 0.5\% of its current 
value ($P_{\rm N}/P_{\rm U}=1.96$). 

Each simulation included one million disk particles distributed from just outside Neptune's initial orbit to 30~AU. 
Their radial profile was set such that the disk surface density $\Sigma \propto 1/r$, where $r$ is the heliocentric 
distance. The large number of disk particles was needed because the capture probability in different parts of the 
Kuiper belt is expected to be $\sim$$10^{-3}$-$10^{-4}$. With $10^6$ disk particles initially, this yields 
$\sim$100-1000 captured particles, and allows us to perform a detailed comparison of the model results with observations 
(Section 4). The disk particles were assumed to be massless such that their gravity does not interfere with the
migration/damping routines. This means that the precession frequencies of planets are not affected by the disk
in our simulations, while in reality they were. This is an important approximation, because the orbital precession 
of Neptune during its high-eccentricity phase can influence the degree of secular excitation of the CCs 
(Batygin et al. 2011). 

All simulations were run to 1 Gyr. The interesting cases were extended to 4 Gyr with the 
standard {\tt swift\_rmvs4} code (i.e., without migration/damping in the 1 to 4 Gyr interval). We performed 
eighteen simulations in total. Three of these runs were designed to test the reproducibility of the results. While 
the results concerning the orbital distribution of bodies implanted into the Kuiper belt (Sections 5.1-5.3) 
were found to be strictly reproducible, the efficiency of capture in the Kuiper belt can vary by a factor of 
a few depending on the behavior of Neptune's eccentricity during the simulation (see discussion in Section 5.4).
  
An additional uncertain parameter concerns the dynamical structure of the original planetesimal disk. It is typically 
assumed that the disk was dynamically cold with orbital eccentricities $e\lesssim0.1$ and 
orbital inclinations $i\lesssim10^\circ$. Some dynamical excitation could have been supplied by scattering off 
of Pluto-sized and larger objects that presumably formed in the disk (Stern \& Colwell 1997, Kenyon et al. 2008).\footnote{The 
escape velocity from Pluto is 1.2 km s$^{-1}$, about 20\% of the Keplerian orbital speed at 25~AU. Therefore, 
eccentricities up to $\simeq$0.2 and inclinations up to $\simeq$12$^\circ$ can be expected from a surface-grazing 
flyby near a Pluto-class object.} The magnitude of the initial excitation is uncertain, because it depends on several 
unknown parameters (e.g., the number of massive objects in the disk). 
Here we operate under the assumption that
the orbital inclinations of disk particles were relatively small initially, and were excited during the main 
stage of planetary instability/migration, when bodies were implanted into the Kuiper belt. This is a reasonable 
assumption, given that the notion of planetary instability/migration was developed, among other reasons,
to explain the complex orbital structure of the Kuiper belt. It would thus seem unsatisfactory to ``resolve'' the 
inclination problem discussed in Section 2 by postulating that the inclinations were already large initially 
(unless it is explained how that happened). The initial eccentricities and initial inclinations of disk 
particles in our standard simulations were distributed according to the Rayleigh distribution with 
$\sigma_e=0.05$ and $\sigma_i=2^\circ$, where $\sigma$ is the usual scale parameter of the Rayleigh distribution
(the mean of the Rayleigh distribution is equal to $\sqrt{\pi/2}\sigma$).
For completeness, we also tested $\sigma_e=0.1$ and $\sigma_e=0.2$, and $\sigma_i=5^\circ$ and $\sigma_i=10^\circ$ 
in several cases.
\section{Constraints and the CFEPS Detection Simulator}
The results of the simulations described in the previous section were compared to observations. We paid  
special attention to the inclination problem described in Section 2, but also made sure that the best models 
identified here are consistent with other Kuiper belt constraints. Our primary constraints were:
\begin{enumerate}
\item The capture efficiency and orbital distribution of HCs. According to Fraser et al. (2014), the HCs 
contain a mass $M_{\rm HC} \simeq 0.01$ $M_{\rm E}$, where $M_{\rm E}=6\times10^{27}$ g is the Earth's mass.
With $M_{\rm disk}=20$ $M_{\rm E}$, the capture probability of HCs would therefore be $P_{\rm HC} \simeq 
0.01/20=5\times10^{-4}$. This estimate is probably at least a factor of $\sim$2 uncertain, because both
$M_{\rm HC}$ and $M_{\rm disk}$ are somewhat uncertain. The inclination distribution of HCs obtained in the 
model is required to be similar to the wide inclination distribution inferred from observations. A detailed 
comparison is done with the Canada-France Ecliptic Plane Survey (CFEPS) detection simulator (see 
below). The distribution of $a$ and $e$ follows a trend seen in Figure \ref{kbos}, where larger 
values of $a$ correspond to larger values of $e$. This trend, which can be an important diagnostic of the 
implantation mechanism, must be reproduced in a successful model. Also, the eccentricities of objects 
captured in the main belt must reach below 0.05, as they do in reality. 
\item The CCs at $42<a<45$ AU must survive and their orbits cannot be excited too much. Dawson \& Murray-Clay 
(2012) suggested that the eccentricities of CCs were {\it not} excited above 0.05 in the inner part of the main belt
($42<a<43.5$ AU) and above 0.1 in the outer part ($43.5<a<45$ AU), because there appears to be a 
stable but unpopulated region above these limits. Morbidelli et al. (2014) demonstrated, however, that the 
7:4 and 9:5 resonances could have depleted the region in question if Neptune was on a somewhat eccentric 
orbit ($e_{\rm N}\simeq0.1$), when it reached $a_{\rm N}=28$~AU, and migrated slowly. Our results, discussed 
in Section 5, are in line with these findings. We do not explicitly discuss the CCs in the following text,
because our main results were obtained with $e_{\rm N} \leq 0.1$. We checked that the CCs are not excessively 
excited in this case, in agreement with Dawson \& Murray-Clay (2012; also see Wolff et al. 2012).
\item The capture efficiency and orbital distribution of the resonant objects. According to the CFEPS survey, 
there are $\sim$3.5 as many HCs as Plutinos with absolute magnitude $H<8$ (diameter $D>150$ km for 0.05 albedo) (B. Gladman, 
personal communication; Petit et al. 2011, Gladman et al. 2012). This suggests a capture probability into 
the 3:2 resonance of $P_{3:2} \simeq 1.5 \times 10^{-4}$ (estimate at least a factor of $\sim$2 uncertain). 
The orbits of Plutinos show moderate to high inclinations, similar to those found for the HCs, and eccentricities 
mainly in the 0.1-0.35 interval. The populations in the 2:1 and 5:2 resonances with Neptune are probably somewhat 
smaller ($\sim$2-4 times) than Plutinos (e.g., Gladman et al. 2014). The population of NTs is much smaller
(Alexandersen et al. 2015), indicating a capture probability of the order of $P_{1:1} \sim 10^{-6}$. 
All resonant populations have a wide inclination distribution (e.g., $\sigma_i>11^\circ$ for NTs; Parker 2015). 
\item The existence and orbits of the detached objects. The detached objects have stable 
non-resonant orbits with semimajor axes beyond Neptune's 2:1 mean motion resonance ($a>47.8$ AU) and
perihelion distances up to $q\simeq40$ AU. These objects cannot be placed on their orbits in the current 
configuration of the planetary orbits and thus provide an important constraint on any formation model. 
Levison et al. (2008) suggested that the detached disk was created during a phase when Neptune had 
a substantial orbital eccentricity ($e_{\rm N}\sim0.3$) and was capable of scattering objects up to 
$q\simeq40$~AU. Here we show that the detached disk can be obtained even for $e_{\rm N} \simeq 0$,
assuming that $\tau \gtrsim 10$~Myr (consistent with the condition required from the inclination constraint). 
The dynamical mechanism responsible for the formation of the detached disk (and HCs) is found to be a 
three-step process related to the capture of scattered bodies in migrating Neptune resonances 
(Section 5.2; Gomes 2003, Gomes et al. 2005, Brasil et al. 2014a). 
\end{enumerate}
We used the CFEPS detection simulator (Kavelaars et al. 2009) to compare the orbital distributions obtained 
in our simulations with observations. CFEPS is one of the largest Kuiper belt surveys with published characterization
(currently 169 objects; Petit et al. 2011). The simulator was developed by the CFEPS team to aid the 
interpretation of their observations. Given intrinsic orbital and magnitude distributions, the CFEPS simulator
returns a sample of objects that would have been detected by the survey, accounting for 
flux biases, pointing history, rate cuts and object leakage (Kavelaars et al. 2009).   
In the present work, we input our model populations in the simulator to compute the detection statistics. 
We then compare the orbital distribution of the detected objects with the actual CFEPS detections using the 
Kolmogorov-Smirnov (K-S) test (Press et al. 1992). 

This is done as follows. The CFEPS simulator takes as an input: (1) the orbital element distribution from 
our numerical model, and (2) an assumed absolute magnitude ($H$) distribution. As for (1), the input orbital 
distribution was produced by a short integration starting from the final model state of the Kuiper belt. 
The orbital elements of each object were recorded at 100 yr intervals during this integration until the 
total number of recorded data points reached $\simeq$$10^5$. Each data point was then treated as an independent 
observational target. We rotated the reference system such that the orbital phase of Neptune in each time 
frame corresponded to its ecliptic coordinates at the epoch of CFEPS observations. This procedure guaranteed 
that the sky positions of the objects in Neptune's resonances were correctly distributed relative to the 
pointing direction of the CFEPS frames.

The magnitude distribution was taken from Fraser et al. (2014). It was assumed to be described by a broken 
power law with 
$N(H)\,{\rm d}H=10^{\alpha_1(H-H_0)}\,{\rm d}H$ for $H<H_{\rm B}$ 
and  
$N(H)\,{\rm d}H=10^{\alpha_2(H-H_0)+(\alpha_1-\alpha_2)(H_{\rm B}-H_0)}\,{\rm d}H$ for $H>H_{\rm B}$,
where $\alpha_1$ and $\alpha_2$ are the power-law slopes for objects brighter and fainter than 
the transition, or break magnitude $H_{\rm B}$, and $H_0$ is a normalization constant.  Fraser et al. (2014) 
found that $\alpha_1=0.9$, $\alpha_2=0.2$ and $H_{\rm B}=8$ for the HCs. In the context of a model where the 
HCs formed at $<30$ AU, and were implanted into the Kuiper belt by size-independent processes (our 
integrations do not have any size-dependent component), the HC magnitude distribution should be shared 
by all populations that originated from $<30$ AU (Morbidelli et al. 2009b, Fraser et al. 2014). We varied 
the parameters of the input magnitude distribution to understand the sensitivity of the results to
various assumptions. We found that small variations of $\alpha_1$, $\alpha_2$ and $H_{\rm B}$ within
the uncertainties given in Fraser et al. (2014) have essentially no effect.    
\section{Results}
\subsection{A Reference Case}
\begin{figure}
\epsscale{0.6}
\plotone{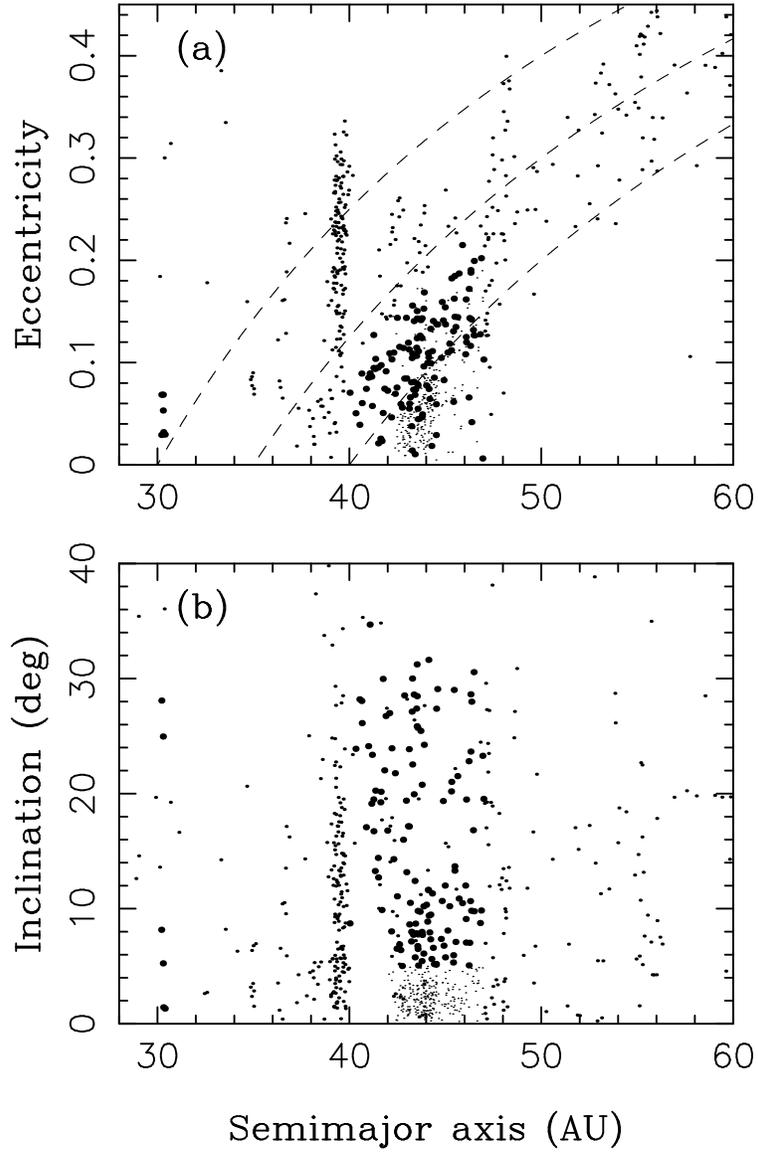}
\caption{The same as Figure \ref{fig1} but without labeling of different populations. This plot is useful for 
a visual comparison with the model results.}
\label{kbos}
\end{figure}

\begin{figure}
\epsscale{0.6}
\plotone{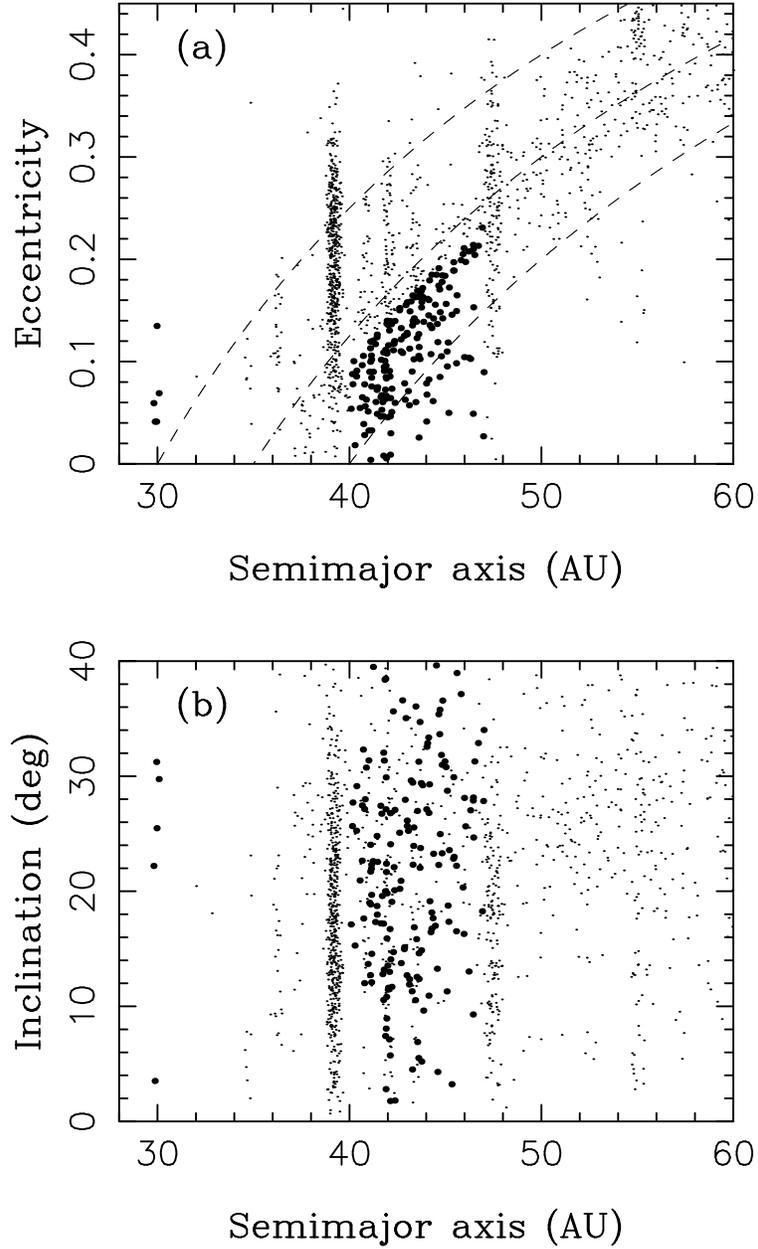}
\caption{The orbital elements of bodies captured in the Kuiper belt in a model with $a_{\rm N,0}=24$~AU
and $\tau=30$~Myr. The HCs and NTs are denoted by larger symbols.}
\label{refsim}
\end{figure}

We first discuss a reference simulation with slow migration of Neptune ($\tau=30$ Myr, $a_{\rm N,0}=24$ AU,
$e_{\rm N,0}=0$, $i_{\rm N,0}=0$), to illustrate that the results of this model match the orbital structure of 
the Kuiper belt, including the wide inclination distribution of the HCs and resonant populations. Later, in 
Section 5.3, we will explain how the results differ from the reference case when various model parameters, 
such as $\tau$ and $a_{\rm N,0}$, are varied.\footnote{These results are strictly reproducible. There is 
enough information given in Section 3 for anyone to repeat our simulations and confirm the results.}
Figure~\ref{refsim} shows the orbital distribution of the model orbits obtained with $a_{\rm N,0}=24$~AU
and $\tau=30$ Myr. This figure can be compared to Figure~\ref{kbos}, but note that 
some caution needs to be exercised in this comparison, because Figure~\ref{kbos} includes various 
observational biases, while Figure~\ref{refsim} does not. Also, the total number of points in the two 
plots is different (known KBOs with good orbits in Figure~\ref{kbos} and a fraction of the initial 
$10^6$ disk particles in Figure~\ref{refsim}). 
 
The model results in Figure~\ref{refsim} show a remarkable similarity to Figure~\ref{kbos}.
The orbital structure obtained in the model shows all main components of the Kuiper belt, including the 
resonant populations, classical belt, scattered and detached disks. The resonances such as the 5:4, 4:3, 
5:3, 7:4 and 5:2 are also populated. The model orbits in the detached disk have perihelion distances 
reaching toward $q\simeq40$ AU, as they do in reality. [Note that the CCs are not shown in 
Figure~\ref{refsim}, because the model discussed here does not account for objects that formed beyond 
30 AU (see discussion of the CCs in Section 1 and the description of the model in Section 3).]

The model distribution of orbital inclinations in Figure \ref{refsim} covers the whole interval between 0 
and 40$^\circ$. This is a notable result because the original disk orbits had $\sigma_i=2^\circ$. The orbital 
inclinations have therefore been significantly excited during the implantation process. The dynamical processes 
responsible for the implantation of objects in the Kuiper belt and their effects on the orbital inclination
are discussed in Section 5.2. Here we first more carefully compare the model distribution with observations.
To do this, the model distribution shown in Figure \ref{refsim} was passed through the CFEPS detection simulator. 
Figure \ref{refsim_ks} shows how the model detections compare with the actual CFEPS detections. 
The comparison is done separately for the 3:2 resonance and HCs. The reason for this is that 
the implantation process and stability properties can, and indeed do, produce differences between these 
populations. We do not show a similar comparison for other resonances, because the number of 
actual CFEPS detections in the 1:1, 2:1 and other resonances is very small (1 to 5 objects detected) and 
a rigorous comparison is therefore not possible at this time. 

Figure \ref{refsim_ks} shows that the orbital distribution of the detected objects in the 3:2 resonance 
agrees with the distribution of the CFEPS detections. The K-S test indicates that the model and observed 
distributions shown in Figures \ref{refsim_ks}a (eccentricity) and \ref{refsim_ks}b (inclination) have 
67\% and 84\% probabilities, respectively, of being derived from the same underlying distribution. This 
is very good agreement.\footnote{To be more precise, a K-S probability of 0.67 (or 0.84) means the 
following. Assume that there was a single parent distribution for both the model and the observations. 
In particular, the model was a random sample containing, say, J entries, while the observations contained 
K entries. If we were to generate two random representations of this parent population, one with J entries 
and one with K, there would be a 67\% (or 84\%) chance that the comparison between these random populations 
would be worse than what we have in Figure \ref{refsim_ks}a (or Figure \ref{refsim_ks}b). 
This holds despite the fact that these new distributions 
were directly derived from the parent. Therefore, the agreement between our model and the observations 
is very good. Indeed, any comparison with a K-S probability greater than $\sim$0.1 should be considered 
acceptable.} The model 
distribution of inclinations obtained with $\tau=30$~Myr is much wider than the one obtained for 
$\tau=1$~Myr (Figure \ref{iprob}), and matches observations very well. The eccentricity distributions
are also very similar. Most Plutinos have $e>0.15$ ($>$80\% of detections). This characteristic 
is a consequence of the implantation mechanism, where orbits are deposited into the 3:2 resonance 
from the scattered disk and retain somewhat large eccentricities (see Section 5.2). Note that,
with $a_{\rm N,0}=24$ AU, the 3:2 resonance is initially outside the outer boundary of the planetesimal 
disk. Plutinos are therefore {\it not} captured in our model from the low-eccentricity orbits as in, 
for example, Hahn \& Malhotra (2005).     

The agreement for the HCs is also good. The K-S test applied to the eccentricity and inclination distributions 
shown in Figures \ref{refsim_ks}c and \ref{refsim_ks}d gives 37\% and 75\% probabilities, respectively. As for 
the inclination distribution of the HCs, our model with $\tau=30$ Myr and $a_{\rm N,0}=24$ AU predicts that about 
10\% of the CFEPS detections should have $i>30^\circ$, while no object was thus far detected by CFEPS with such 
a high inclination. This is not a problem, however, because the CFEPS detected only 10 HCs with $i>10^\circ$, and 
the statistical constraints for $i>30^\circ$ are very weak. For $i<10^\circ$, on the other hand, there are 
concerns with contamination from the CCs, which are not modeled here. The CCs should clearly have the 
dominant contribution to the statistics for $i<5^\circ$. For $5^\circ<i<10^\circ$, the situation is unclear. 
There are 11 CFEPS detections in this intermediate inclination range, which is similar to the number of detections 
for $i>10^\circ$. In our model with $\tau=30$ Myr, however, these intermediate inclinations are not populated 
as much (we get about half of the expected detections). We believe this issue arises because our model with 
smooth migration of Neptune and a fixed value of $\tau$ is only an approximation of the real evolution
(see discussion in Section 6).
\subsection{The Implantation Mechanism}
We examined the orbital histories of test particles in the reference simulation and found that the implantation 
of bodies from $<30$ AU into the Kuiper belt is in general a three-step process. The first two steps are common 
for the HCs and resonant populations; the third step is what distinguishes them. We first describe these 
steps and then illustrate them with a few examples. Specifically:    
\begin{description}
\item[Step 1:] The disk planetesimals are scattered by Neptune from $<$30 AU to $>$30 AU. Their distribution
resembles that of the scattered disk in that they populate the region with $a>30$ AU and $q \lesssim 
a_{\rm N}(t)$, where $a_{\rm N}(t)$ is the semimajor axis of migrating Neptune. Here we define the Intermediate 
Source Region or ISR as the orbital region with $q \leq Q_{\rm N}(t)$, where $Q_{\rm N}(t)=a_{\rm N}(1+e_{\rm N})$ 
is Neptune's aphelion distance, and $40<a<47$~AU. Most bodies implanted into the main belt evolved onto their 
present orbits via the ISR.
\item[Step 2:] The scattered bodies evolve onto orbits with large libration amplitudes in mean motion 
resonances. The secular dynamics inside the mean motion resonances is complex, including large-amplitude 
Kozai, apsidal and nodal cycles (e.g., Morbidelli et al. 1995, Nesvorn\'y \& Roig 2000, 2001). These effects 
can act to decrease the orbital eccentricity, thus decoupling the orbit from Neptune, on a characteristic 
timescale that is comparable to the period of the secular oscillations ($>$1 Myr). 
\item[Step 3:] If Neptune were not migrating, the evolution described in Step 2 would be reversible and bodies 
would be released, sooner or later, back to the scattered disk. With Neptune's migration, however, two additional 
alternatives can happen: (1) the orbit can evolve to a smaller libration amplitude and stabilize inside the 
resonance, or (2) it can be released from the resonance with low eccentricity and can end up on a stable, HC-like orbit 
with $q>35$ AU.     
\end{description}
The three-step implantation mechanism described above was originally proposed in Gomes (2003) and Gomes et 
al. (2005). We will therefore call it the {\it Gomes mechanism} in the following.\footnote{The Gomes mechanism was 
previously shown to work in several self-consistent simulations of the planetary instability/migration, where 
the disk particles carried actual mass. Because a relatively small number of disk particles was used, however, 
Neptune's migration was unrealistically grainy in these simulations. Here we show that the Gomes mechanism works, 
with a somewhat lower efficiency, even if Neptune's migration was smooth.} 
 
Figure \ref{tp1} shows an example of a disk particle that was captured on a high-inclination and 
low-eccentricity orbit in the main belt. This case is a clear illustration of the three-step Gomes mechanism 
described above. Initially, the particle starts with $a=28.5$ AU, $e=0.04$ and $i=3.5^\circ$. It is scattered 
by Neptune and evolves into the scattered disk, where it remains until roughly $t=13$ Myr after the start of the 
simulation. During this first stage, the eccentricity and inclination are excited by encounters with
Neptune. The orbit is then captured in the 2:1 resonance and remains in the resonance until $t=22$ Myr
(see the libration of the resonant argument in panel f). Once captured, it undergoes Kozai oscillations
(see the libration of the perihelion argument in panel e). The eccentricity drops from 0.4 to 0.06 
during this phase (13 to 22 Myr), while the inclination further increases from 18$^\circ$ to 28$^\circ$.
Finally, about 22 Myr after the start of the simulation, the particle is released from the 2:1 resonance
(see panels b and f) and lands on a main-belt orbit with $a=42.8$ AU, $e=0.05$ and $i=28^\circ$. This orbit 
is stable for 4 Gyr.     

Figure \ref{tp2} shows an example of a disk particle that was captured on a high-inclination orbit 
in the 3:2 resonance. In this case, the scattering phase lasted until about $t=65$ Myr. Both the eccentricity 
and inclination were strongly excited during this phase. At $t\simeq65$~Myr, the orbit entered into the 3:2
resonance. The resonant libration amplitude was initially variable but later evolved to become $\simeq$100$^\circ$
(panel f).
The 3:2 resonant orbits with such amplitudes are stable (Nesvorn\'y \& Roig 2000). Indeed, the orbit stayed in the 
3:2 resonance for the whole duration of our integration (4 Gyr). During the capture into the 3:2 resonance 
the orbit started showing Kozai cycles ($\omega$ started librating in panel e while $e$ and $i$ in panels 
c and d exhibited signs of correlated oscillations). The Kozai oscillations with (full) amplitude of about 80$^\circ$ 
remained for the whole duration of the simulation. This case is reminiscent of Pluto, whose orbit also 
has Kozai cycles, but the final orbital inclination of this particle is considerably higher
($i=36^\circ$ vs. Pluto's $i\simeq17^\circ$).    

Figure \ref{tp3} is a case of Neptune Trojan. Akin to the example of the captured Plutino discussed 
above, the scattering phase lasts very long ($\simeq62$ Myr). The eccentricity is excited but the inclination 
remains relatively low ($<$15$^\circ$). Then, around $t=50$ Myr, the orbit starts showing signs
of the Kozai resonance (panel e) and undergoes two brief periods during which the 1:1 resonant
angle librates (about $t=54$ Myr and $t=60$ Myr). The inclination rises and eccentricity drops in 
an anti-correlated pattern. The orbit is caught into the 1:1 resonance at $t\simeq62$~Myr, 
eventually stabilizes with a very small libration amplitude ($\simeq30^\circ$) around the leading
Lagrangian point $L_4$, and remains there for 4 Gyr. 

The three examples discussed above were selected from hundreds of similar cases to illustrate the Gomes 
mechanism for capture in the main belt and resonances, and the high inclinations that these orbits
can reach, if the migration of Neptune is slow and the scattering phase lasts long. We found that the 2:1
resonance was typically involved in capture of orbits in the main belt, but resonances such as 
7:4 or 9:5 also contributed. The orbits of the detached objects produced in our simulations follow a 
similar pattern, but the resonances responsible for raising their perihelia are different 
(e.g., 5:2, 7:3, 3:1). Moreover, relatively strong resonances such as the 3:1 are capable 
of producing the detached orbits with very large perihelion distances, and some of these orbits are 
reminiscent of that of 2004 XR$_{190}$ (Buffy) ($a=57.7$ AU, $q=51.5$ AU; Allen et al. 2006; see 
Figure \ref{refsim}; also Gomes 2011).     
\subsection{The Inclination Distributions Obtained in Different Models}
The results of our simulations show that the inclination distribution of bodies implanted in the Kuiper
belt depends both on $\tau$ and $a_{\rm N,0}$, but the main effect is that of $\tau$. This is because the 
implantation of bodies in the Kuiper belt happens on a timescale comparable to $\tau$. With short 
$\tau$, the implantation must be fast and Neptune does not have much time to raise the inclinations of
the scattered bodies. Consequently, the inclination distribution of the implanted bodies is narrow and 
clustered toward $i\sim0$. If, on the other hand, $\tau$ is long, the orbital inclinations of scattered 
objects can be substantially excited by Neptune encounters before these bodies are implanted into 
the Kuiper belt. The inclination distribution of the implanted bodies is thus wide in this case. 
Figures \ref{iprob} and \ref{refsim_ks} illustrated the dependence on $\tau$ for $\tau=1$ Myr and 
$\tau=30$ Myr. Figure \ref{itau} show this dependence for several additional cases.  

The inclination distribution of Plutinos obtained with different migration timescales is shown in Figure 
\ref{itau}a. The results obtained for $\tau<10$ Myr and any $a_{\rm N,0}$ are clearly a poor fit, because 
they indicate $\sigma_i \lesssim 5^\circ$, while Plutinos have $\sigma_i>10^\circ$ (Kavelaars et al. 2008, 
Gulbis et al. 2010, Gladman et al. 2012). The best-fit distribution for the case with $\tau=10$~Myr  
and $a_{\rm N,0}=24$~AU is obtained for $\sigma_i \simeq 10^\circ$. When this case is compared to the CFEPS 
via the CFEPS detection simulator (Figure \ref{10My_ks}b), we find that it can be ruled out with 99.6\% 
confidence. The results with $\tau=10$ Myr and $a_{\rm N,0}>24$~AU can be ruled out at even higher confidence 
levels, because the inclination distributions of Plutinos obtained in those cases are narrower than the 
one obtained with $\tau=10$ Myr and $a_{\rm N,0}=24$~AU. We conclude that fast migration timescales 
with $\tau\lesssim10$ Myr do not work for Plutinos.

Longer timescales produce better results. The case with $\tau=30$ Myr and $a_{\rm N,0}=24$~AU was discussed 
in Section 5.1 and clearly matches observations very well (Figure \ref{refsim_ks}). Interestingly, the 
inclination distribution obtained in the model also depends on $a_{\rm N,0}$. For example, the inclination 
distributions obtained with $\tau=30$ Myr and $a_{\rm N,0}=26$~AU or $a_{\rm N,0}=28$~AU are much narrower 
($\sigma_i \simeq 10^\circ$) than the one obtained with $\tau=30$ Myr and $a_{\rm N,0}=24$~AU ($\sigma_i=15$
-20$^\circ$). When compared to the CFEPS detections, these cases can be ruled out at a $>$99\% confidence level. 
This happens because bodies tend to be captured relatively early in these simulations when Neptune's migration rate 
is still substantial. In contrast, the long-range migration with $a_{\rm N,0}\lesssim25$~AU offers more 
opportunity for capture at late times, and therefore leads to a wider inclination distribution that is more 
in line with observations. We therefore conclude from the inclination distribution of Plutinos that Neptune's 
migration was slow ($\tau\gtrsim30$ Myr) and long range ($a_{\rm N,0}\lesssim25$~AU).

An additional argument that favors a long migration timescale comes from the eccentricity distribution of
Plutinos. With $\tau\leq10$ Myr, the model eccentricity distribution is skewed toward large values
of $e$ when compared to observations. This can be demonstrated by comparing the model to the CFEPS
detections via the CFEPS detection simulator. For example, for $\tau=10$ Myr and $a_{\rm N,0}=24$~AU
(Figure \ref{10My_ks}a), the K-S test applied to the eccentricity distributions of the detected 
Plutinos gives only a 0.3\% probability that the two distributions are the same. This mismatch is 
a consequence of step~2 of the capture process, discussed in Section 5.2, where there is not enough 
time available with rapid migration to decrease eccentricities. The eccentricities of 
captured objects therefore end up being too large.

We now move to discussing the HC inclination distribution. The HCs are deposited into the main belt 
($40<a<47$ AU) by mean motion resonances such as the 2:1 and 7:4. With $a_{\rm N,0}=28$ AU, these resonances 
are located in the main belt with $a=44.4$ AU and $a=40.7$~AU, respectively. Therefore, in this case, 
the implantation of HCs into the main belt begins almost immediately after the start of a simulation 
(with a short delay required for Neptune to scatter bodies to the ISR; Section 5.2). The bodies 
implanted into the main belt during the initial stages will have small orbital inclinations and will 
skew the inclination distribution of HCs toward small values. Thus, even if long migration
timescales are used in this case (e.g., $\tau=30$ Myr or 100 Myr), the inclination distributions 
obtained in the model end up being incorrect. See, for example, the case with $a_{\rm N,0}=28$ AU and 
$\tau=30$~Myr in Figure \ref{itau}b. This specific case indicates $\sigma_i=6$-10$^\circ$; it can be 
ruled out at a 99.8\% confidence level from the CFEPS detections. 

For $a_{\rm N,0}=24$ AU, on the other hand, the 2:1 and 7:4 resonances are at 38.1 and 34.9~AU. These 
resonances therefore {\it cannot} deposit bodies into the main belt during the initial stages of 
migration (because they are not located in the main belt during these initial stages).
This means that the orbital inclinations in the ISR region can be excited by Neptune's encounters 
{\it before} the main phase of the implantation starts. It starts when the 2:1 resonance 
moves beyond $\simeq$40~AU, or equivalently, when Neptune moves beyond $a\simeq40/2^{2/3}=25.2$ AU. 
When exactly this happens depends both on $a_{\rm N,0}$ and $\tau$. For example, with $a_{\rm N,0}=24$~AU 
and $\tau=30$ Myr, Neptune moves past 25.2~AU at $t\simeq10$ Myr after the start of migration. 
This delay is long enough to raise the mean orbital inclination in the ISR region to $\simeq$10$^\circ$. 
This explains why the orbital inclinations of HCs are generally higher when $a_{\rm N,0}$ is smaller. 

Figure \ref{10My_ks} shows the eccentricity (panel c) and inclination (panel d) distributions of HCs 
obtained for $a_{\rm N,0}=24$ AU and $\tau=10$ Myr. While the eccentricity distribution is (slightly) 
discrepant when compared to the CFEPS detections, the inclination distribution looks good (K-S 
probability 29\%). Unlike in Figure \ref{refsim_ks}d, panel d of Figure \ref{10My_ks} compares the 
inclination distributions all the way down to $5^\circ$. The CFEPS inclination distribution is steep 
from $5^\circ$ to $10^\circ$, and shallow above $10^\circ$. This may suggest that the underlying 
distribution is bimodal, perhaps because it resulted from capture at two different stages of 
Neptune's migration (see discussion in Section 6). The model distribution obtained with $a_{\rm N,0}=24$ 
AU and $\tau=10$~Myr is a good proxy for the overall shape of the CFEPS curve. We conclude 
from the inclination distribution of HCs that Neptune's migration was slow ($\tau\gtrsim10$ Myr) 
and long range ($a_{\rm N,0}\lesssim25$~AU).

Figures \ref{isr}a shows the mean orbital inclination of bodies in the ISR as a function of time. The mean 
inclination of the ISR population steadily increases with time. 
It is $\simeq$5$^\circ$ at $10^6$ yr, $\simeq$10$^\circ$ at $10^7$ yr, $\simeq$15$^\circ$ at $3\times10^7$ yr, 
and $\simeq$20$^\circ$ at $10^8$ yr. This makes it obvious that bodies captured in the main belt late will 
have, on average, larger orbital inclinations than bodies captured early, and explains the trends 
discussed above (see also Figure \ref{idistr}). Note, however, that the number of bodies available in the 
ISR drops for $t>10$ Myr (Figure \ref{isr}b). This is because bodies in the scattered disk evolve to very 
long orbital periods, or move to short orbital periods and are subsequently ejected from the solar system 
by Jupiter. This means that the number of bodies available for capture at very late times is relatively 
small. The very late captures ($t>10^8$ yr) are therefore not very important for the overall statistics.

In summary, we find that the inclination distribution of the HCs is a reflection of the inclination 
distribution in the ISR, weighted by the number of bodies in the ISR, and time integrated over the 
capture window that depends both on $a_{\rm N,0}$ and $\tau$. In addition, the inclination distribution 
becomes modified by the dynamical processes involved in step 2 of the capture process (see Section 5.2). 
Our main conclusion from the inclination constraint is that Neptune's migration was long-range 
($a_{\rm N,0}\lesssim25$~AU), and that the migration timescale was long ($\tau\gtrsim10$~Myr). The case with 
$a_{\rm N,0}=24$~AU and $\tau=10$~Myr works relatively well for the HCs, but it fails for Plutinos (Figure 
\ref{10My_ks}). The case with $a_{\rm N,0}=24$~AU and $\tau=30$~Myr works for HCs with $i>10$ deg {\it and} 
Plutinos. As we discuss in Section~6, the somewhat different timescales indicated by the HCs and Plutinos 
may be related to a two-stage migration of Neptune, with faster migration ($\tau\simeq10$~Myr) during the 
first stage and slower migration ($\tau\simeq30$~Myr) during the second. 
\subsection{The Implantation Efficiency}
The efficiency of implantation of the disk bodies into the Kuiper belt is a product of partial efficiencies of 
the three steps described in Section 5.2. During the first step, bodies from the disk at $<30$ AU are scattered 
by Neptune to $>30$ AU, where they can be captured into resonances. The number of scattered bodies available 
for capture is a function of time (Figure \ref{isr}b). The capture efficiency in a resonance mainly depends 
on the resonance strength (e.g., the strong 3:2 resonance is expected to capture more bodies) and Neptune's 
migration speed. In step 2, the resonant bodies can evolve to orbits with lower eccentricities assuming there 
is enough time for the secular cycles to act. This depends on how Neptune's migration timescale compares with
the period of secular cycles in a specific resonance. Also, since the secular cycles depend on Neptune's 
eccentricity, this stage is affected by Neptune's eccentricity behavior during migration. Finally, whether a 
resonant orbit is or is not released from a resonance during step 3, is mainly influenced by 
Neptune's migration speed (more bodies are released for higher speeds). Moreover, the implantation of bodies 
into the main belt occurs only when the relevant resonances are present in the $40<a<47$~AU region. For 
implantation via the 2:1 resonance, this requires that Neptune is beyond $\simeq$25 AU.

Some of the trends can be identified in our results. For example, with $a_{\rm N,0}=24$ AU, the efficiency 
of implantation on a stable orbit in the 3:2 resonance is $P_{\rm 3:2}=9.2\times10^{-4}$ for $\tau=10$ Myr,
$P_{\rm 3:2}=5.3\times10^{-4}$ for $\tau=30$ Myr, and $P_{\rm 3:2}=2.0\times10^{-4}$ for $\tau=100$ 
Myr. Also, for $\tau=30$ Myr, $P_{\rm 3:2}=5.3\times10^{-4}$ for $a_{\rm N,0}=24$ AU, $P_{\rm 3:2}=1.2\times10^{-3}$ 
for $a_{\rm N,0}=26$~AU, and $P_{\rm 3:2}=2.3\times10^{-3}$ for $a_{\rm N,0}=28$ AU. Thus, a longer migration 
timescale leads to lower $P_{\rm 3:2}$, and larger $a_{\rm N,0}$ leads to higher $P_{\rm 3:2}$. The trends for 
the implantation in the HC region are similar. For example, with $\tau=30$ Myr, $P_{\rm HC}=1.9\times10^{-4}$ 
for $a_{\rm N,0}=24$ AU, $P_{\rm HC}=2.1\times10^{-4}$ for $a_{\rm N,0}=26$ AU, and $P_{\rm HC}=1.1\times10^{-3}$ 
for $a_{\rm N,0}=28$ AU.\footnote{$P_{\rm 3:2}$ (or $P_{\rm HC}$) is the probability that an original disk object 
ends up on a stable orbit in the 3:2 resonance (or in the HC region). We compute this probability by dividing 
the number of 3:2 resonant (or HC) bodies at the end of our simulations (4 Gyr) by the number of bodies 
in the original disk ($10^6$).}

In our preferred case ($a_{\rm N,0}=24$ AU and $\tau=10$ or 30 Myr), the main-belt capture efficiency 
is $P_{\rm HC}\simeq2$-$4\times10^{-4}$ for each initial particle in the original disk. With 
$M_{\rm disk}=20$~$M_{\rm Earth}$, the total mass of the hot population would therefore be $M_{\rm HC}=0.004$-0.008~$M_{\rm Earth}$.
This is satisfactory when compared to  $M_{\rm HC} \simeq 0.01$ $M_{\rm E}$ estimated by Fraser et al. 
(2014), especially because Fraser's estimate has a considerable uncertainty. Also, scaling from Jupiter 
Trojans, there should have been (3-$4)\times10^7$ planetesimals with absolute magnitude $H<9$ in the original 
disk (Nesvorn\'y et al. 2013). With $P_{\rm HC}\simeq2$-$4\times10^{-4}$, our model would predict $\sim$7000-14,000 
HCs with $H<9$, while Adams et al. (2014) give $19,000\pm5,000$ from the Deep Ecliptic Survey (DES) for the 
whole population of the classical belt. This agreement is reasonable. [We note that the model estimates discussed 
here were obtained with $e_{\rm N,0}=0$ and whenever $e_{\rm N}$ stayed low during the migration.
The cases with $e_{\rm N,0}=0.1$ and/or the ones where the mean motion resonances between Uranus and Neptune 
acted to temporarly increase $e_{\rm N}$ during the migration tend to produce larger $P_{\rm HC}$, by a factor of 
a few, but more simulations would need to be done to establish this trend convincingly. The trend could be 
related to the dependence of the secular cycles inside the mean motion resonances on $e_{\rm N,0}$.]
 
A major problem is identified when we consider the capture probability of the {\it resonant} objects. With 
$P_{\rm 3:2}\simeq5\times10^{-4}$ for the preferred case with $a_{\rm N,0}=24$ AU and $\tau=30$ Myr, we would 
predict that the 3:2 resonance population should host $\sim1.5$-3 times more objects than the hot population
in the main belt, while according to the CFEPS survey there are $\sim$3.5 times as many HCs as Plutinos (with 
absolute magnitude $H<8$; B. Gladman, personal communication). This would indicate $P_{\rm 3:2}/P_{\rm HC}\simeq0.3$
(this estimate has a $<$50\% formal uncertainty). The 3:2 resonance is thus obviously overpopulated in our 
simulations, roughly by a factor of 5-10. We call this the {\it resonance overpopulation problem}. This problem 
was already noted in many previous dynamical models of Kuiper belt formation (e.g., Hahn \& Malhotra 2005, 
Levison et al. 2008, Morbidelli et al. 2008). 

There are several potential solutions to this problem. For example, we performed several simulations 
with $\tau = 100$ Myr and found that $P_{\rm HC}$ tends to be higher, and $P_{\rm 3:2}$ tends to be lower,
than in the cases with $\tau = 30$ My. For example, with $a_{\rm N,0}=26$ AU and $\tau=100$~Myr (see Figure 
\ref{tau100} for the orbital distribution of bodies obtained in this simulation), we found that
$P_{\rm HC}\simeq 9 \times 10^{-4}$ and $P_{\rm 3:2}=1.6 \times 10^{-4}$, thus indicating 
$P_{\rm 3:2}/P_{\rm HC} \simeq 5.6$. Also, with $a_{\rm N,0}=28$ AU and $\tau=100$ Myr, $P_{\rm HC}\simeq 3.2 \times 
10^{-3}$ and $P_{\rm 3:2}=1.4 \times 10^{-3}$, so $P_{\rm 3:2}/P_{\rm HC} \simeq 2.3$. These ratios are
more similar to the value $P_{\rm 3:2}/P_{\rm HC}\sim 3.5$ inferred from observations. The case with 
$a_{\rm N,0}=26$~AU produces a slightly wider inclination distribution than indicated by observations,
while the one with $a_{\rm N,0}=26$ AU produces a much narrower inclination distribution, thus suggesting the possibility that 
an intermediate value of $a_{\rm N,0}\simeq26$-27 AU would give the correct result. We do not give much emphasis
to the cases with $\tau=100$ Myr in this paper, because it is not clear how these cases relate to
our current models of the planetary instability and migration (e.g., they may require a very low mass of the 
planetesimal disk). New planetary instability simulations, with a focus on very slow migration timescales, 
will need to be performed.  Another solution of the resonance overpopulation problem, which will also require 
additional modeling effort that goes beyond the scope of this paper, is discussed in Section 6. 

We point out that the overpopulation problem is 
not specific to the 3:2 resonance. Instead, nearly all resonances are overpopulated. On the other hand, when we compare,
relative to each other, the number of bodies captured in different resonances in our simulations, we find 
that these populations have roughly the right proportions. For example, for $a_{\rm N,0}=24$ AU and $\tau=30$ Myr, 
we find that $P_{\rm 3:2}/P_{\rm 2:1}\simeq3$ and $P_{\rm 3:2}/P_{\rm 2:1}\simeq7$. This is comparable to the resonance  
population statistics in the 3:2, 2:1 and 5:2 resonances discussed in Gladman et al. (2014) (even though 
the population in the 5:2 resonance was previously thought to be comparable to that in the 3:2 resonance; 
Gladman et al. 2012). 

Another notable result obtained from our simulations concerns the NTs. Two problems were identified 
in previous modeling of the NT capture: (1) the capture efficiency obtained in the previous simulations 
was $\sim$2 orders of magnitude too high, and (2) the inclination distribution was too narrow (e.g., Nesvorn\'y \& 
Vokrouhlick\'y 2009, Parker 2015). Related to (1), only five NTs were captured out of the original $10^6$ disk 
particles for $a_{\rm N,0}=24$~AU and $\tau=30$ Myr. This indicates the capture probability 
$P_{\rm NT}\sim5\times10^{-6}$ and is more in line with observations (Alexandersen et al. 2014). As for (2), 
four of five stable NTs produced by the reference run have inclinations $>20^\circ$ (Figure \ref{refsim}). 
This is encouraging, but better statistics will be needed to compare things more carefully.  
\section{Discussion and Conclusions}
The Gomes mechanism was identified here to have fundamental importance for the origin of dynamical
structure in the Kuiper belt. The basic requirement for the Gomes mechanism to work is that the migration
timescale $\tau$ is comparable to, or longer than, the secular cycles inside the mean motion resonances
such as 2:1, 3:2, 7:3, etc. With $a_{\rm N,0}\lesssim25$~AU, the 2:1 resonance is initially below 40 AU,
sweeps over the main belt location at $40<a<47$ AU during Neptune's migration, and is responsible for
the delivery of most objects into the main belt. Since the secular cycles in the 2:1 resonance have 
a several-Myr period, $\tau$ needs to be at least several Myr for this to work. If, instead, $\tau\sim1$ 
Myr, objects can still be captured in the main belt region, assuming that Neptune had an orbit with substantial 
orbital eccentricity (Levison et al. 2008). In this case, orbits evolve from the scattered disk to the main 
belt by normal secular cycles outside the mean motion resonances (Dawson \& Murray-Clay 2012). These cycles 
have a shorter period and work for shorter migration timescales.

A major problem with the high-eccentricity ($e_{\rm N}>0.1$) phase of Neptune is the opposing constraints from 
the hot and cold populations, as explained in Dawson \& Murray-Clay (2012).
Specifically, to preserve the CCs, the eccentricity of Neptune cannot be large and/or must be damped 
fast (Batygin et al. 2011, Wolff et al. 2012). To capture the HCs by the normal secular cycles 
outside the mean motion resonances, however, Neptune's initial eccentricity must be relatively high
at some point during Neptune's migration. While these two constraints rule out most of parameter space, 
Dawson \& Murray-Clay (2012), working under the assumption that the HCs were captured by the normal secular 
cycles, found solutions that satisfy both. This niche of parameter space is $e_{\rm N,0}\simeq 0.1$ and 
$a_{\rm N,0} > 28$~AU, essentially meaning that Neptune's migration would have to be short range.

The beauty of the Gomes mechanism is that it works even if Neptune's eccentricity was never large. This is 
because this mechanism does not rely on the normal secular eccentricity oscillations forced by eccentric Neptune on 
orbits outside the mean motion resonances. Instead, it appears as a product of large eccentricity 
oscillations due to the existence of large-amplitude secular cycles inside the mean motion resonances, akin 
to those first pointed out for the 3:1 Jupiter resonance in the asteroid main belt (Wisdom 1982). Here we showed that the 
Gomes mechanism is the dominant implantation mechanism from $<30$~AU in a regime where $e_{\rm N}$ is low. 
Therefore, the argument of Dawson \& Murray-Clay (2012) that $e_{\rm N,0}>0.12$ is required to explain the HCs 
does not apply. On a related note, Levison et al. (2008) suggested that the existence of the detached disk with orbital 
perihelia extending to $\simeq$40~AU is a consequence of an eccentric phase of Neptune, for Neptune to be 
capable of scattering objects to $q\simeq40$ AU, and used $e_{\rm N}=0.3$ in their simulations. Here we 
showed that the Gomes mechanism can produce the correct orbital architecture of the detached disk even 
if $e_{\rm N}$ stays low, assuming that the migration timescale was long (e.g., Figures~4 and 13).

In this work, we stressed the importance of the inclination distribution of the KBOs. This is because 
the inclination distribution has been relatively well characterized from observations and can 
therefore be used to constrain models. We showed that the Gomes mechanism is capable of producing the 
observed wide inclination distribution from a dynamically cold disk at $<$30 AU, assuming that Neptune's
migration was long-range ($a_{\rm N,0}\lesssim25$) AU and slow ($\tau\gtrsim10$ Myr).\footnote{Using 
an initially strongly excited disk is counterproductive, because this has the consequence, as shown by our additional 
simulations of Neptune migrating into a pre-heated disk, that the implantation efficiency in the main belt 
drops by a factor of several.} Since the Gomes mechanism is insensitive to Neptune's eccentricity, the 
eccentricity could have been negligible during Neptune's migration in much the same way as originally proposed by 
Malhotra (1993, 1995) and later used by Hahn \& Malhotra (1999, 2005) to model the origin of the Kuiper 
belt from a disk at $>30$ AU.

Does this mean that the Nice-type instability never happened? Not really. On one hand, the results presented
here seem to rule out the strong instability version of the Nice model, where Neptune was thrown onto an eccentric 
orbit with $a>25$ AU. This is because, as we discussed above, the short-range migration of Neptune into a 
dynamically cold disk at $<30$ AU would lead to a narrow distribution of orbital inclinations, in contradiction
to observations (also see Levison et al. 2008). 
Moreover, to stabilize eccentric Neptune at $>25$ AU, the disk would have to be massive 
($\simeq50$ $M_{\rm Earth}$; Nesvorn\'y \& Morbidelli 2012; hereafter NM12), and would produce fast migration of 
Neptune, thus leading to a double contradiction, because both $\tau$ and $a_{\rm N,0}$ would be out of the 
plausible range identified here (Neptune's migration needs to be slow, not fast, to explain the inclination 
distribution). 

On the other hand, some dynamical instability in the outer solar system clearly must have happened. The best evidence 
for this is the eccentric orbit of Jupiter, which can be conveniently explained if Jupiter suffered encounters 
with an ice giant (Morbidelli et al. 2009a). A discontinuous evolution of Jupiter's semimajor axis, known as the 
{\it jumping-Jupiter model}, presumably produced by various scattering events during the epoch of planetary 
encounters, is also required from the terrestrial planet (Brasser et al. 2009, 2013; Agnor \& Lin 2012) and 
asteroid belt constraints (e.g., Morbidelli et al. 2010). Moreover, planetary encounters may be needed to explain 
the capture and orbital distribution of Jupiter Trojans and irregular satellites (Nesvorn\'y et al. 2007, 2013, 2014). 
 
A new model of planetary instability has recently been proposed (Nesvorn\'y 2011, NM12, Batygin et al. 2012). 
This model has been the framework of several newer publications discussed above. It postulates that the 
early solar system had an extra ice giant, which was ejected into interstellar space during the instability. 
Figure \ref{case9} illustrates this possibility. In this model, five outer planets start in a relatively 
relaxed configuration with Neptune at $\simeq22$ AU. The first thing that happens in the simulation 
is that Neptune migrates into the outer disk located at 24-30 AU. After Neptune reaches $\simeq$ 28 AU, the
instability happens, during which the extra ice giant has encounters with all the other outer planets,
and is subsequently ejected by Jupiter. The main features of this model which are most relevant for
the Kuiper belt are: 
(1) Neptune's eccentricity and inclination are never large ($e<0.1$ and $i<2^\circ$), 
(2) the initial mass of the outer disk at $<30$ AU is relatively small ($\simeq15$-20~$M_{\rm Earth}$, NM12), 
therefore implying a slow migration of Neptune,
(3) Neptune's semimajor axis discontinuously changes (by $\simeq$0.2-0.5 AU) 
when Neptune is at $\simeq$28 AU, as a result of one or two very close encounters with the ejected ice giant; 
the migration rate is slower after the jump than it was during the previous migration stage, and 
(4) the ejected ice giant briefly overlaps with the Kuiper belt (Batygin et al. 2012). 

As for (4), we carefully looked into several instability cases from NM12 and found that the relevant period 
during which the ice giant's orbit overlaps with the Kuiper belt is too brief to significantly affect the orbits 
in the Kuiper belt. Batygin et al. (2012), who found larger effects in about 50\% of studied cases, did so 
probably because their work covered a broad range of possibilities, with some of their instability cases 
being somewhat too cataclysmic, in our opinion, to produce the solar system as we know it now. Items (1) and 
(2) present the right conditions for the Gomes mechanism to work and play a dominant role over other 
implantation mechanisms. 

Unlike in the idealized case studied here, Neptune's migration in Figure~\ref{case9}, and other cases reported 
in NM12, happens in two stages. During the first stage, that is, before the instability happens, Neptune 
migrates with $\tau\simeq10$ Myr for $M_{\rm disk}=20$~$M_{\rm Earth}$ or $\tau\simeq20$~Myr for 
$M_{\rm disk}=15$ $M_{\rm Earth}$ (these are the best exponential fits in the NM12 cases we looked at).
During the second stage, that is after the instability, Neptune migrates with $\tau\simeq30$~Myr for 
$M_{\rm disk}=20$ $M_{\rm Earth}$ or $\tau\simeq50$ Myr for $M_{\rm disk}=15$ $M_{\rm Earth}$. These best-fit 
$\tau$ values are only approximate, because the real migration is not exactly exponential, and the effective 
$\tau$ is typically longer as time progresses.  

These timescales, and the long-range nature of Neptune's migration in NM12, agree quite nicely with the 
constraints on Neptune's migration derived from the Kuiper belt in this work. This shows that the NM12 instability 
model, which was developed entirely from constraints unrelated to the Kuiper belt, may have some relevance for 
the early evolution of the solar system. 

The inclination distribution of the HCs may provide some evidence for the two stage migration of Neptune 
in the NM12 model. This is because the HCs can be captured into the main belt 
both during the first phase, when Neptune's migration was faster, and during the second phase, when the migration 
was slower. The main implication of this is that the HCs can be a composite of two populations captured at 
two different stages. From the discussion in this paper, these populations are expected to have different
inclination distributions, thus potentially explaining why the CFEPS detections of the HCs show different slopes 
for $5^\circ<i<10^\circ$ (captures during the first stage) and $i>10^\circ$ (second stage). Also, Plutinos and 
other resonant populations, captured during the first stage, would be released when Neptune jumped during the 
NM12 instability. This could relieve the resonance overpopulation problem discussed in Section 5.4. The present 
resonant populations would then have to be captured entirely during the second phase, when the migration of 
Neptune was slower. This could explain why we are seeing a preference in our results for slightly longer $\tau$ 
values for Plutinos than for HCs (e.g., the case with $\tau=10$ Myr is works well for the HCs but does not 
really work for Plutinos; Figure \ref{10My_ks}).  
    
Much work has yet to be done to fully understand the dynamics of the KBOs during Neptune's migration, and 
how the history of Neptune's orbit is constrained by the dynamical structure of the Kuiper belt. 
Clearly, the idealized migration model studied here is a major simplification. We used it to highlight 
several interesting results that can be obtained within the framework of this model. We now plan to increase 
the realism of the model by considering the two-stage migration from NM12. We believe that this will be a 
crucial step toward resolving the resonance overpopulation problem that plagued previous studies of  
Kuiper belt formation. In a companion paper, Nesvorn\'y (2015), we study the implications of the NM12 
instability model for the cold classical belt.    

\acknowledgments
This work was supported by NASA's Outer Planet Research (OPR) program. All CPU-intensive simulations in this 
work were performed on NASA's Pleiades Supercomputer.\footnote{http://www.nas.nasa.gov/hecc/resources/pleiades.html} 
We thank W. F. Bottke, L. Dones, B. Gladman, H. F. Levison, A. Morbidelli, D. Vokrouhlick\'y, and an 
anonymous reviewer for helpful comments on this work.

\clearpage

\clearpage

\clearpage

\clearpage

\clearpage
\begin{figure}
\epsscale{0.7}
\plotone{fig5a.eps}\\[0.7cm]
\plotone{fig5b.eps}
\caption{The cumulative distribution of eccentricities (left) and inclinations (right)
for Plutinos (upper) and HCs (lower). The dashed lines show the actual CFEPS detections (29 Plutinos, and
10 HCs with $i>10^\circ$). The solid lines show the distributions of model bodies ($a_{\rm N,0}=24$~AU
and $\tau=30$~Myr) detected by the CFEPS 
simulator. Both the observed and model distributions plotted here therefore contain the CFEPS observational 
bias. For the HCs, we compare the distributions for $i>10^\circ$ to avoid any potential contamination 
of the detection statistics from the CCs, which are not modeled here.}
\label{refsim_ks}
\end{figure}

\clearpage
\begin{figure}
\epsscale{0.8}
\plotone{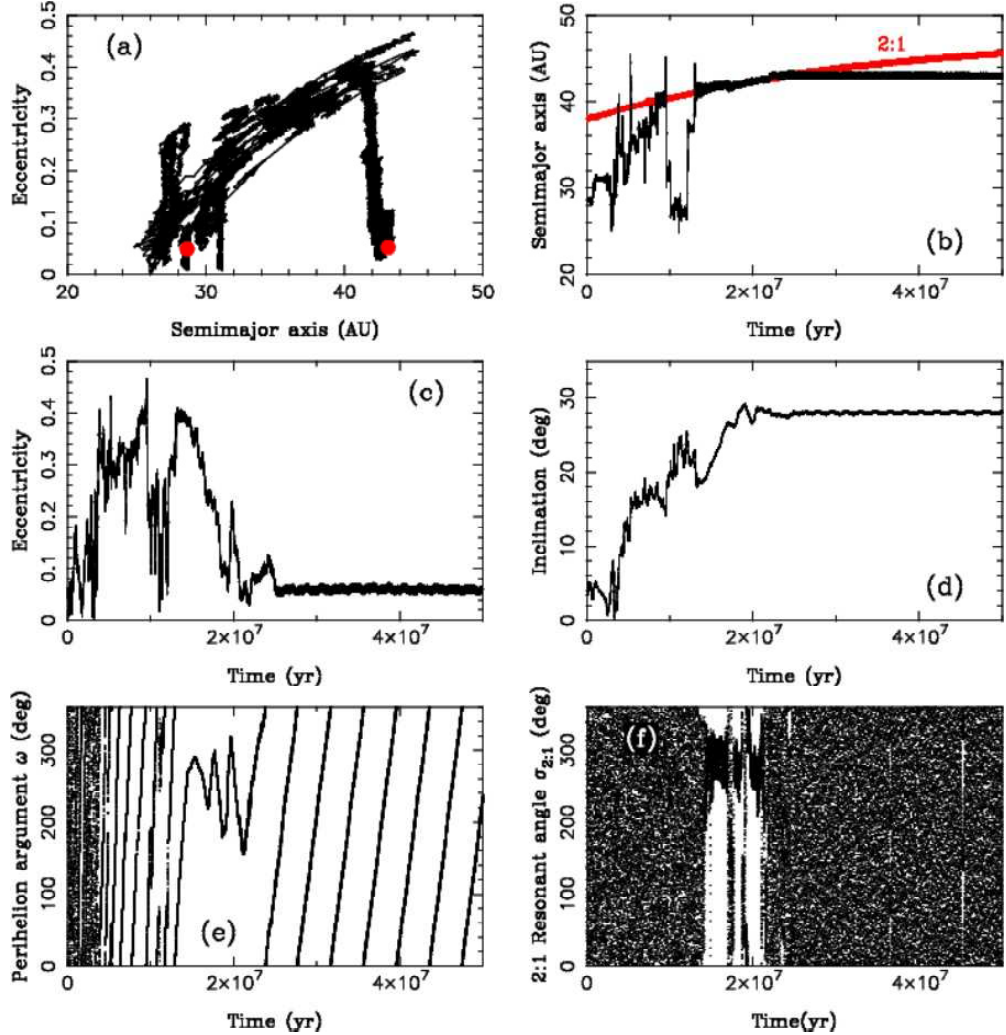}
\caption{An illustration of the Gomes implantation mechanism. The panels show: the (a) path of a disk
particle in the $(a,e)$ projection; the two red dots show the initial and final orbits, (b) semimajor axis,
(c) eccentricity, (d) inclination, (e) perihelion argument $\omega$, and (f) 2:1 resonant angle
$\sigma_{2:1}=2\lambda-\lambda_{\rm N}-\varpi$, where $\lambda$ and $\lambda_{\rm N}$ are the particle's and 
Neptune's mean longitudes, and $\varpi$ is the particle's perihelion longitude. After being scattered by 
Neptune and experiencing Kozai cycles inside the 2:1 mean motion resonance, the disk particle ends up 
on a high-inclination orbit in the main belt.}
\label{tp1}
\end{figure}

\clearpage
\begin{figure}
\epsscale{0.8}
\plotone{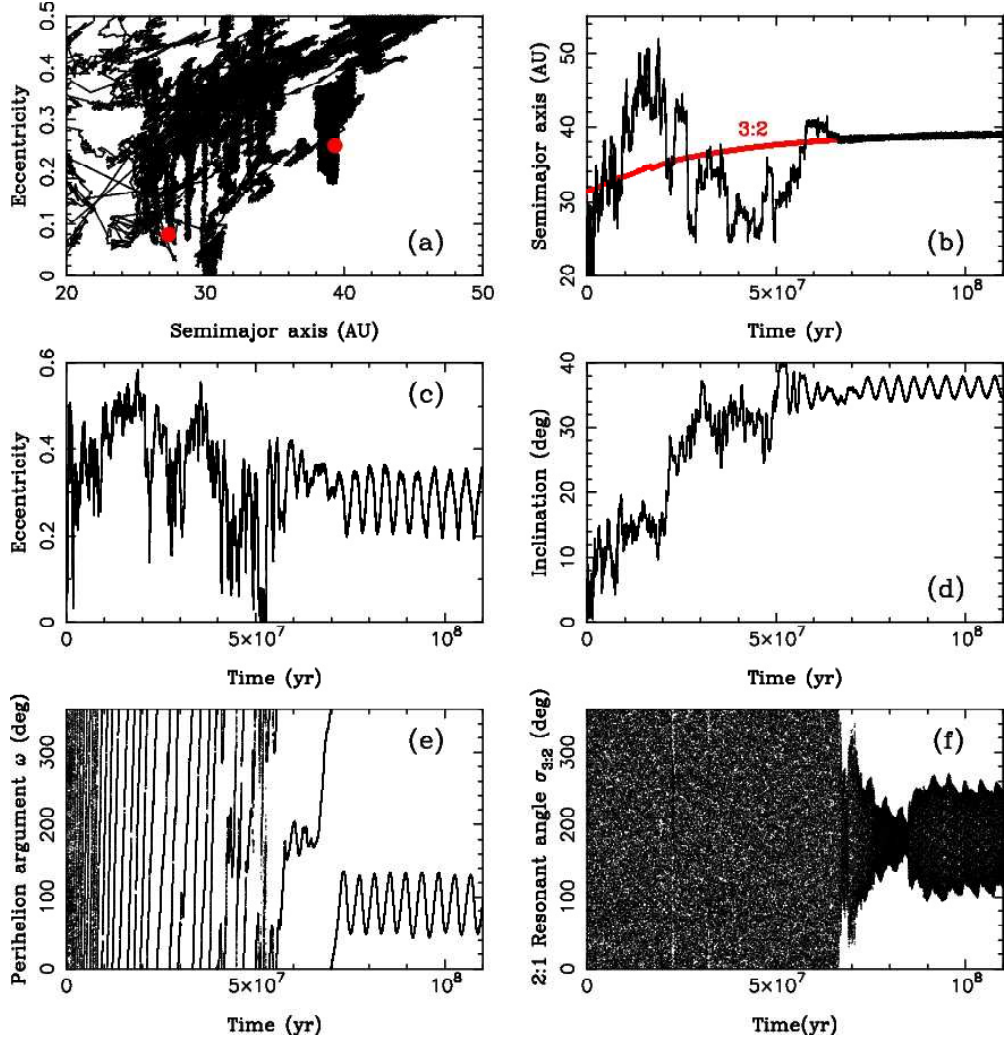}
\caption{Capture of a disk particle on a stable, high-inclination orbit in the 3:2 resonance. 
The panels show: the (a) path of the disk particle in the $(a,e)$ projection; the two red dots show the initial 
and final orbits, (b) semimajor axis, (c) eccentricity, (d) inclination, (e) perihelion argument $\omega$, and 
(f) 3:2 resonant angle $\sigma_{3:2}=3\lambda-2\lambda_{\rm N}-\varpi$.}
\label{tp2}
\end{figure}

\clearpage
\begin{figure}
\epsscale{0.8}
\plotone{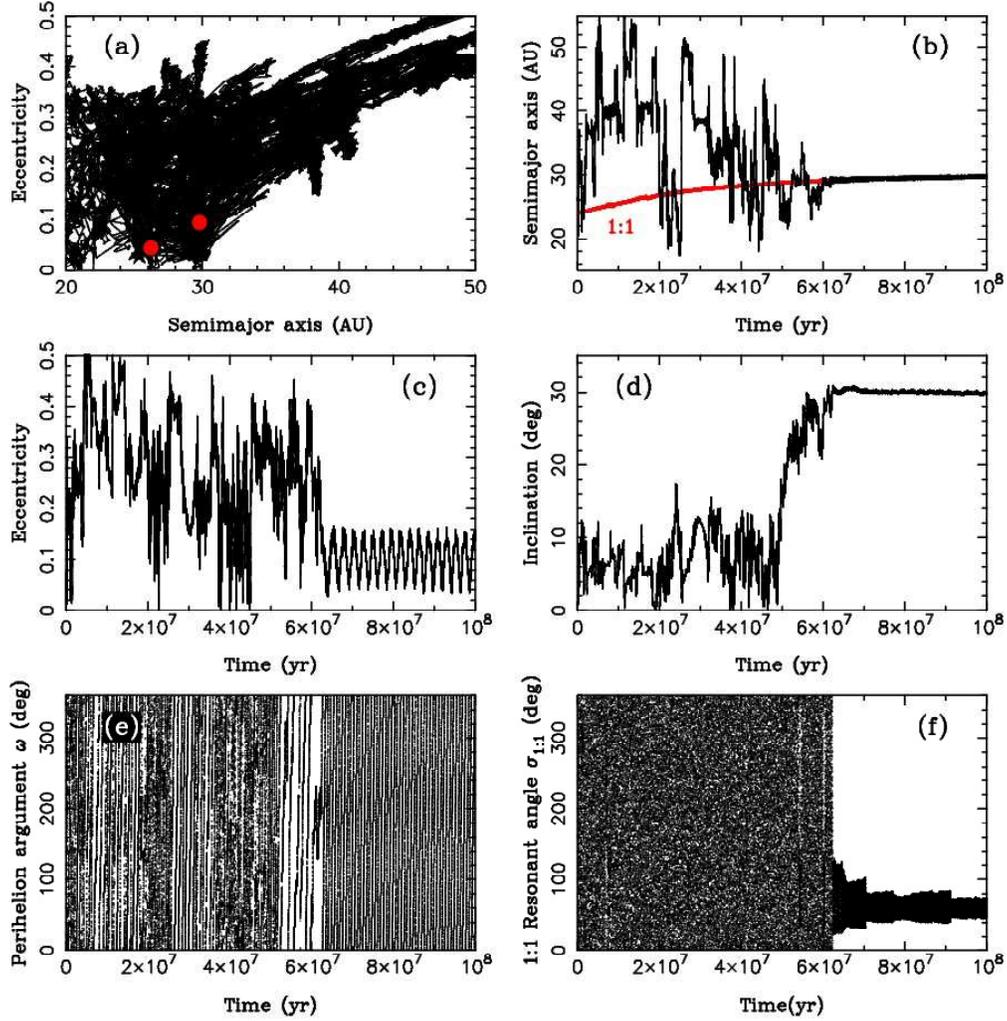}
\caption{Capture of a disk particle on a high-inclination Neptune Trojan orbit. 
The panels show: the (a) path of the disk particle in the $(a,e)$ projection; the two red dots show the initial 
and final orbits, (b) semimajor axis, (c) eccentricity, (d) inclination, (e) perihelion argument $\omega$, and 
(f) 1:1 resonant angle $\sigma_{1:1}=\lambda-\lambda_{\rm N}$.}
\label{tp3}
\end{figure}

\clearpage
\begin{figure}
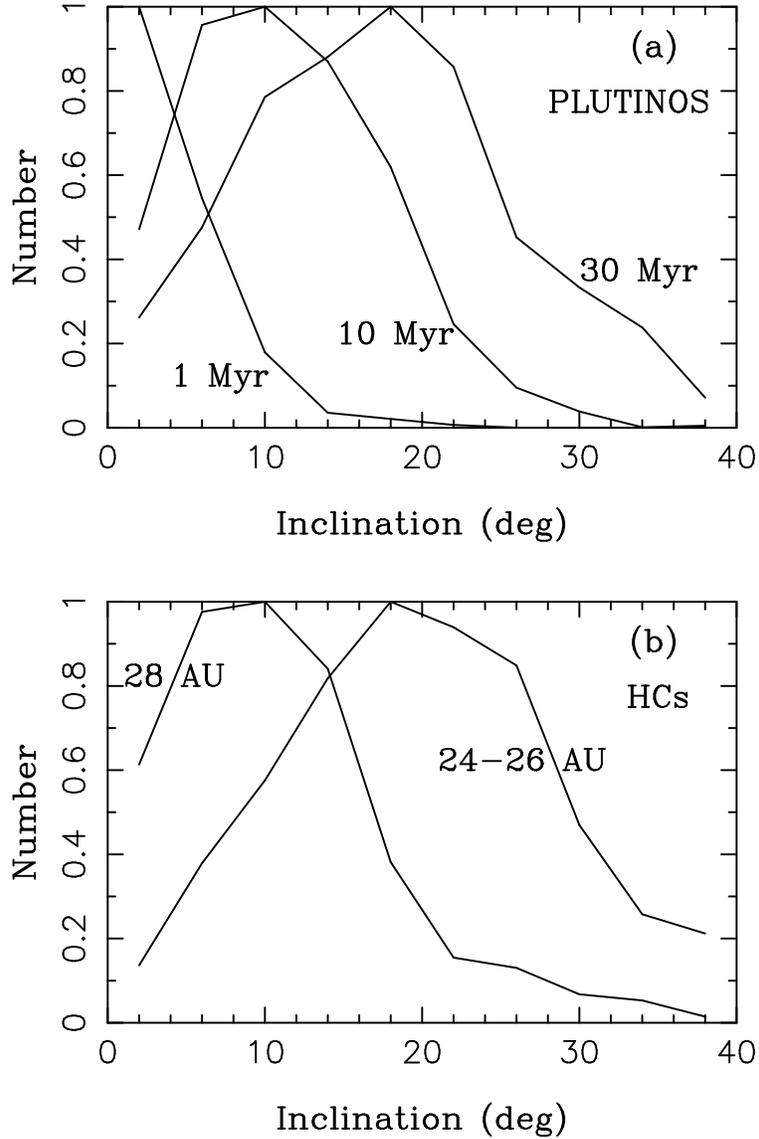

\epsscale{0.6}
\plotone{fig9a.eps}\\[0.7cm]
\plotone{fig9b.eps}
\caption{The dependence of the inclination distribution on $\tau$ and $a_{\rm N,0}$: (a) Plutinos, and 
(b) HCs. These distributions are the {\it intrinsic} distributions obtained in the model (i.e., they
do {\it not} include any observational bias). The inclination distribution of Plutinos in panel (a) is 
sensitive to the assumed migration timescale. The labels in panel (a) denote the distributions obtained for 
different values of $\tau$. In panel (b), we show the inclination distribution of HCs obtained with 
$\tau=30$~Myr. The two lines correspond to different starting positions of Neptune ($a_{\rm N,0}=24$-26 
AU and 28 AU; the cases with $a_{\rm N,0}=24$ and 26 AU were similar and were put together in this plot).}
\label{itau}
\end{figure}

\clearpage
\begin{figure}
\epsscale{0.7}
\plotone{fig10a.eps}\\[0.7cm]
\plotone{fig10b.eps}
\caption{The cumulative distributions of eccentricities (left) and inclinations (right)
for Plutinos (upper) and HCs (lower). The dashed lines show the actual CFEPS detections (29 Plutinos, and
21 HCs with $i>5^\circ$). The solid lines show the distributions of model bodies ($a_{\rm N,0}=24$~AU
and $\tau=10$~Myr) detected by the CFEPS simulator. Both the observed and model distributions plotted 
here therefore contain the observational bias of CFEPS. For the HCs we compare the distributions for 
$i>5^\circ$ to avoid contamination of the detection statistics from the CCs, which are not modeled here.}
\label{10My_ks}
\end{figure}

\clearpage
\begin{figure}
\epsscale{0.6}
\plotone{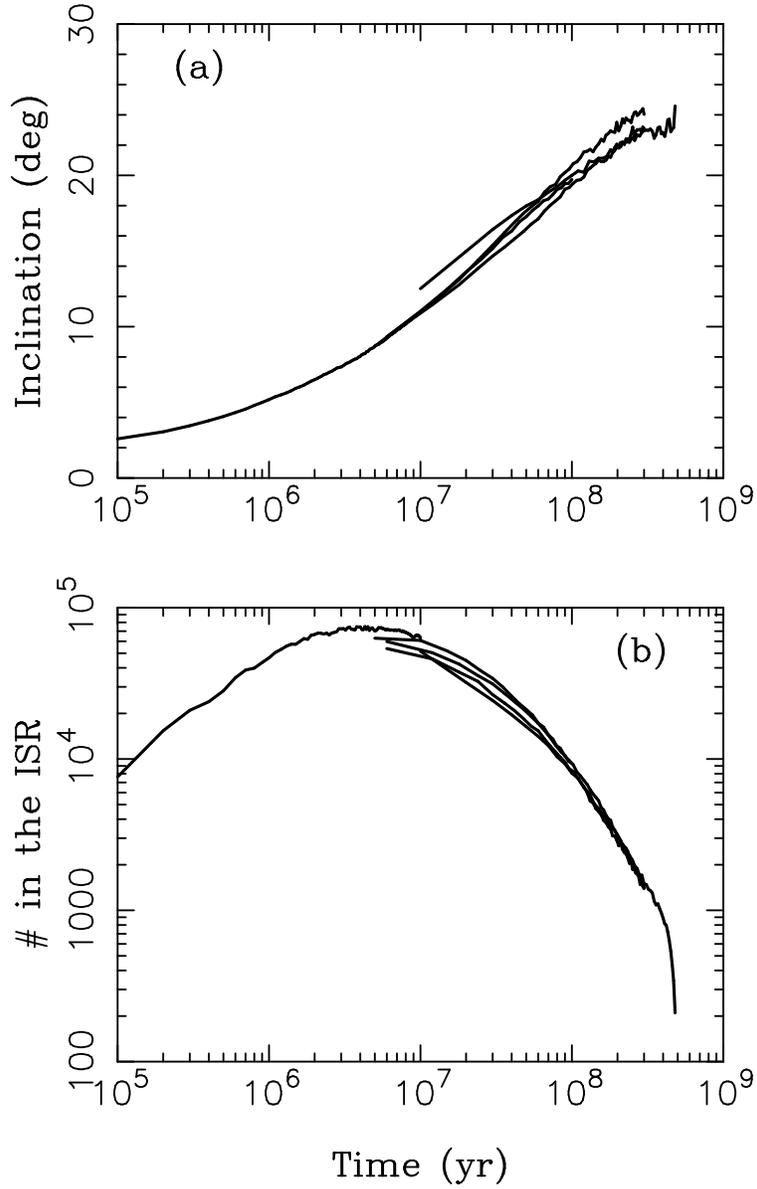}
\caption{The mean orbital inclination and number of objects in the Intermediate Source Region (ISR), defined as
$40<a<47$ AU and $q<Q_{\rm N}$. The lower panel shows that it takes $10^6$ to $10^7$ yr to build up the
ISR population, which then decays by about an order of magnitude over the next $10^8$ years. The mean 
orbital inclination of the ISR population in panel (b) steadily increases with time. The different lines
in panels (a) and (b) correspond to models with different $\tau$ and $a_{\rm N,0}$. The overall shape of the 
lines is insensitive to these parameters.}
\label{isr}
\end{figure}

\clearpage
\begin{figure}
\epsscale{0.8}
\plotone{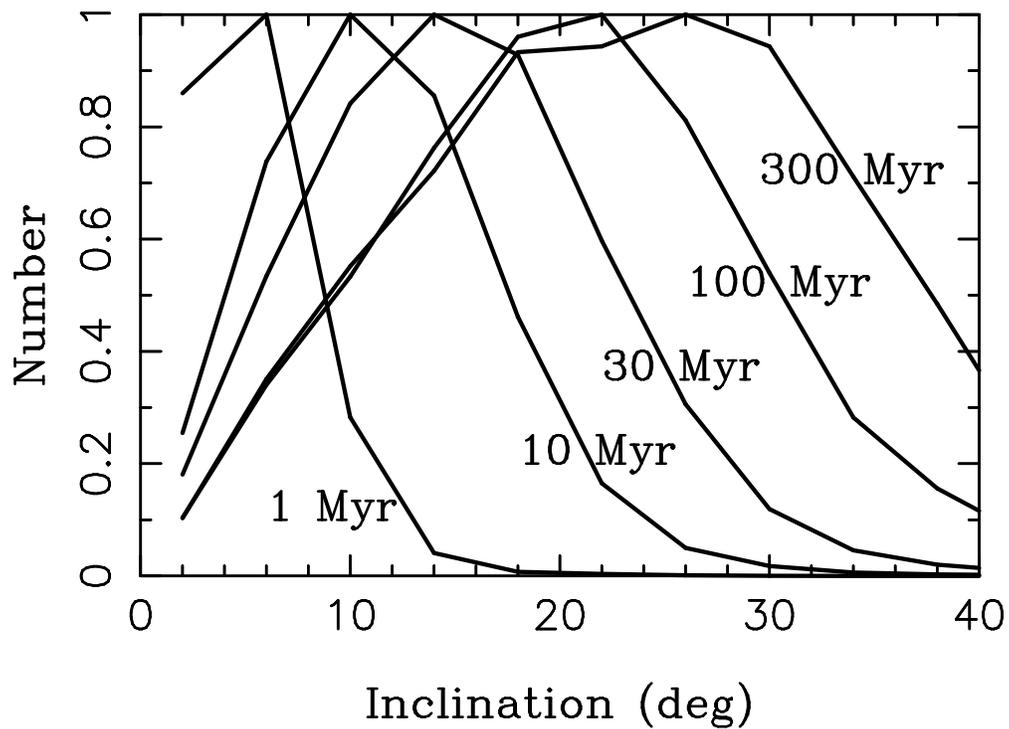}
\caption{The inclination distribution of objects in the IRS for $t=1$, 10, 30, 100, 
and 300 Myr after the start of Neptune's migration. The plot shows how the orbital inclinations in the ISR 
increase as a result of encounters with Neptune. Here we used $a_{\rm N,0}=24$~AU and $\tau=30$~Myr.}
\label{idistr}
\end{figure}

\clearpage
\begin{figure}
\epsscale{0.6}
\plotone{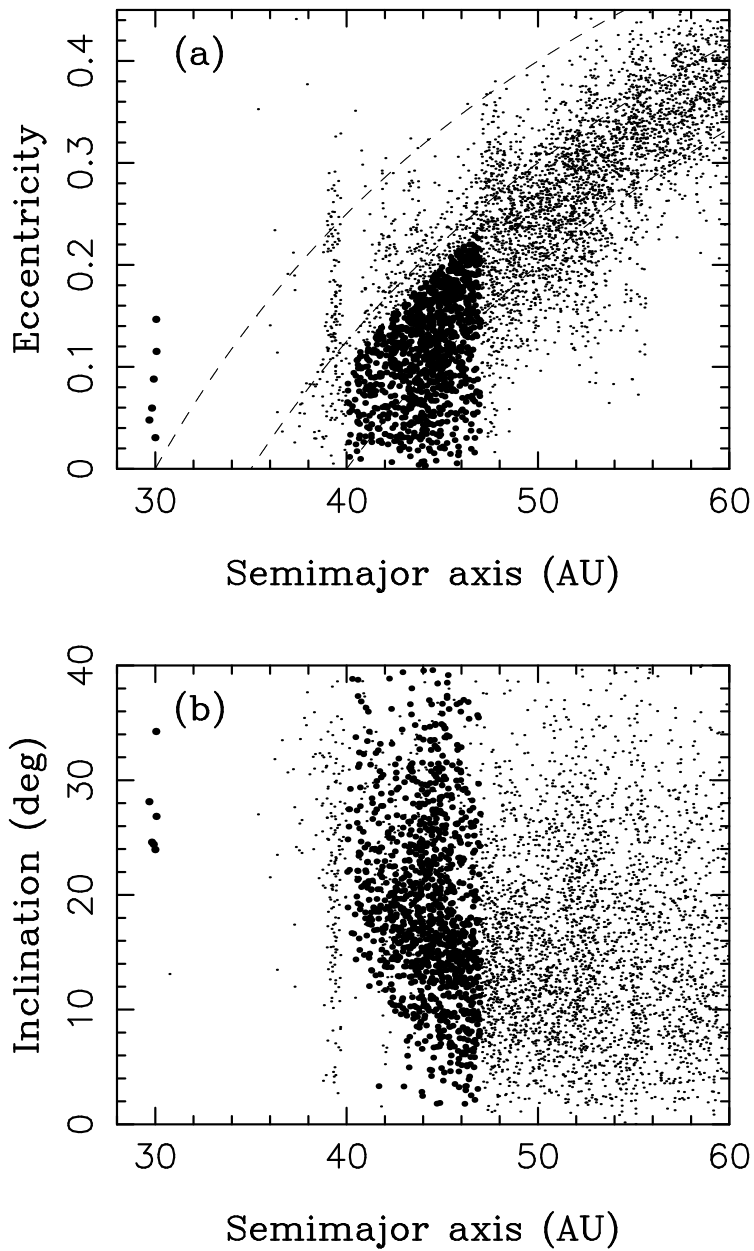}
\caption{The orbital elements of bodies captured in the Kuiper belt in a model with $a_{\rm N,0}=26$~AU
and $\tau=100$~Myr. The HCs and NTs are denoted by larger symbols. Note that the number of bodies 
captured in the HC region is $\simeq$5 times larger than in Figure \ref{refsim}. The detached disk 
beyond 50 AU is also more populated and extends to $q>40$ AU. Both these results are a consequence
of the very slow migration of Neptune assumed in this simulation.}
\label{tau100}
\end{figure}

\clearpage
\begin{figure}
\epsscale{0.9}
\plotone{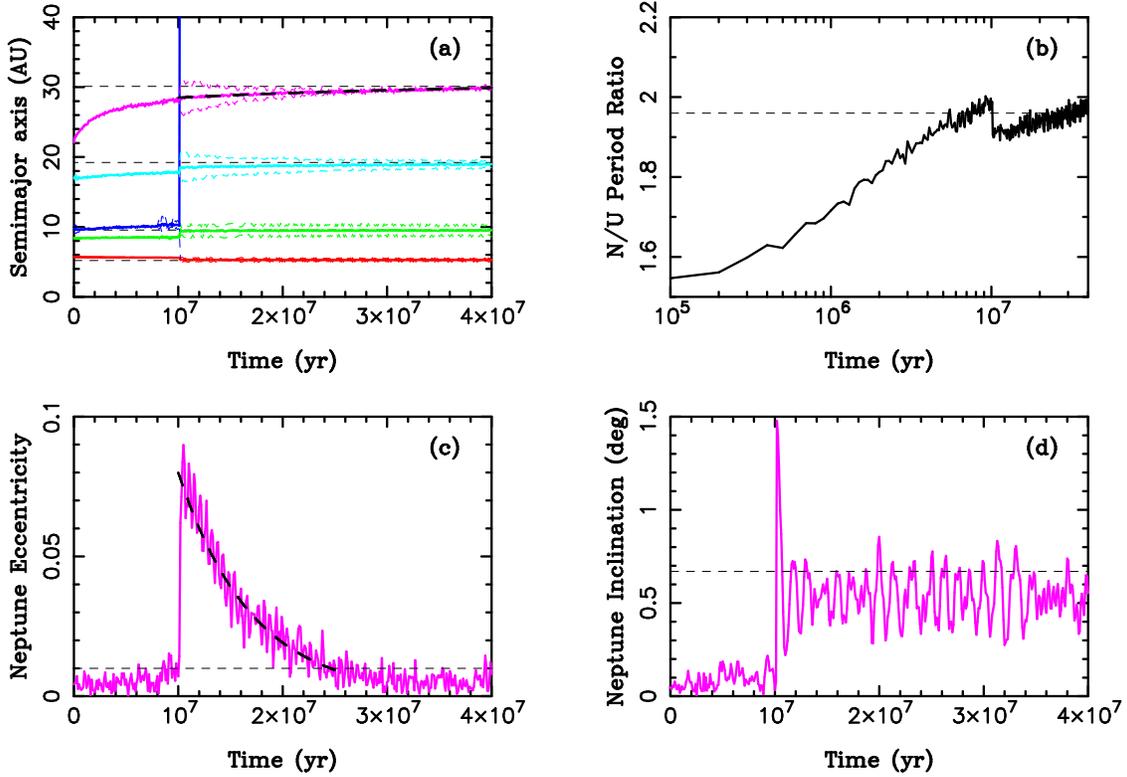}
\caption{The orbital histories of the giant planets from the instability/migration model of NM12. 
In this example, the fifth giant planet was initially placed 
on an orbit between Saturn and Uranus and was given a mass equal to Neptune's mass. The three inner 
planets were started in the (3:2,4:3) resonant chain, and Uranus and Neptune on non-resonant orbits 
with $a_{\rm U}=17$ AU and $a_{\rm N}=22$ AU. Ten thousand particles, representing the outer planetesimal disk, 
were distributed with semimajor axes $23.5<a<29$ AU, surface density $\Sigma=1/a$, and 
low eccentricities and low inclinations. With the total disk mass $M_{\rm disk}=20$ $M_{\rm E}$, each disk particle 
has approximately one Pluto mass. (a) The semimajor axes (solid lines), and perihelion and aphelion distances 
(thin dashed lines) of each planet's orbit. The inner ice giant was ejected into interstellar space at 
$t\simeq10$ Myr after the start of the simulation. The final orbits of the four remaining planets are a good 
match to those in the present solar system (thin dashed lines). (b) The period ratio $P_{\rm N}/P_{\rm U}$. The 
thin dashed line shows $P_{\rm N}/P_{\rm U}=1.96$ corresponding to the present orbits of Uranus and Neptune. 
Panels (c) and (d) show the eccentricity and inclination of Neptune's orbit.  The bold dashed lines in panels 
(a) and (c) approximate Neptune's migration with $\tau_a=30$ Myr and damping of Neptune's eccentricity 
with $\tau_e=7$ Myr. These long migration/damping timescales are characteristic for the migration/instability
models from NM12.}
\label{case9}
\end{figure}

\end{document}